\documentclass[fleqn,usenatbib]{mnras}
\usepackage{graphicx}
\usepackage{float}
\usepackage{wrapfig}
\usepackage{amsmath}
\usepackage{amssymb}
\usepackage{bm}
\usepackage[font=small,labelfont=bf]{caption}
\usepackage{citesort}
\usepackage{threeparttable}
\usepackage{comment}
\usepackage{mathtools}
\usepackage{array}
\usepackage[dvipsnames]{xcolor}
\usepackage[textwidth=1in]{todonotes}
\usepackage{multicol, cuted}
\usepackage[normalem]{ulem}  %

\newcommand{\HI}{H\,{\sc i}}

\newcommand{\cdd}{\mathcal{F}}
\newcommand{\Meudon}{{\sc Meudon PDR}}
\newcommand{\kms}{km\,s$^{-1}$}
\newcommand{\cmsq}{cm$^{-2}$}

\title[OH in the diffuse ISM]{OH in the diffuse interstellar medium: physical modelling and prospects with upcoming SKA precursor/pathfinder surveys}

\author[S. A. Balashev et al.]{
S. A. Balashev,$^{1}$\thanks{E-mail: s.balashev@gmail.com}
N. Gupta,$^2$
D. N. Kosenko,$^{1}$
\\
$^{1}$Ioffe Institute, 26 Politeknicheskaya st., St.\ Petersburg, 194021, Russia\\
$^{2}$ Inter-University Centre for Astronomy and Astrophysics, Post Bag 4, Ganeshkhind, 411 007, Pune, India \\
}

\date{Submitted 21 December 2020}

\pubyear{2020}

\begin{document}

\pagerange{\pageref{firstpage}--\pageref{lastpage}}
\maketitle

\begin{abstract}

Hydroxyl ($\rm OH$) is known to form efficiently in cold gas ($T\sim 100$\,K) along with the molecule $\rm H_2$ and can be used as an efficient tracer of the diffuse molecular gas in the interstellar medium (ISM). 
Using a simple formalism describing the $\rm H I/H_2$ transition and a reduced network of major chemical reactions, we present a semi-analytical prescription to estimate the abundances of O-bearing molecules in the diffuse ISM. 
We show that predictions based on our analytical prescription are in good agreement with the estimates obtained using the MEUDON PDR code which utilizes the full reaction network. 
We investigate the dependence of the relative abundances of OH/H~{\sc i} and $\rm OH/H_2$ on the variations of physical conditions i.e., the metallicity, number density ($n$), cosmic ray ionization rate ($\zeta$) and strength of UV field ($\chi$) in the medium.  
We find that the $\rm OH/H I$ abundances observed in the Galactic ISM can be reproduced by models with $n\sim 50$\,cm$^{-3}$, $\chi\sim 1$ (Mathis field) and $\zeta\sim3\times10^{-17}$\,s$^{-1}$, with a variation of about 1\,dex allowed around these values. 
Using the constrained $\rm H_2$ column density distribution function at $z\sim3$, we estimate the $\rm OH$ column density distribution function and discuss future prospects with the upcoming large radio absorption line surveys.

\end{abstract} 


\section{Introduction}
\label{firstpage}

The cold atomic (\HI) and molecular ($\rm H_2$) phases of the interstellar medium (ISM) are basic fuel for star formation in galaxies. For this reason, understanding the physical properties and evolution of these two phases is of great interest and crucial to understand many key observables -- such as the cosmic evolution of star formation rate density \citep[SFRD;][]{Madau2014} -- related to the galaxy evolution. In a diffuse molecular cloud, these two phases are intimately linked to each other through an important ingredient of the ISM physics -- the \HI/$\rm H_2$ transition. This transition occurs in a very narrow region of the medium with typical temperatures, $T\sim 100$\,K and number densities, $n\sim 100$\,cm$^{-3}$.

The cold atomic phase which is classically referred to as the cold neutral medium \citep[CNM;][]{Heiles2003a} can be easily observed through \HI\ 21-cm line in emission and absorption.  However, due to the weakness of the transition the emission lines are mostly detected in the Galaxy and the local Universe. In turn, the major constituent of the molecular ISM i.e., H$_2$ is extremely hard to observe in emission.  
Consequently, it is mostly studied indirectly, through other molecular species such as $\rm CO$ and $\rm HCO^{+}$, whose production is coupled with the presence of substantial amount of $\rm H_2$ in the medium. 
\citep{Dame2001, Leroy2009, Tacconi2018, Freundlich2019}. However, these molecular emission line observations mostly trace the dense molecular gas, which generally represent clumps of the cold ISM embedded in the elongated envelopes of diffuse gas. In the local Universe, the diffuse molecular gas can be detected via far-UV Lyman- and Werner-band absorption lines of $\rm H_2$. 
At $z\gtrsim2$ these lines are redshifted to optical wavelengths, which has enabled sensitive observations of diffuse molecular gas in distant galaxies using large ground-based telescopes.
It has been recently established, that this diffuse molecular gas, that is not traced by CO \citep{Balashev2017}, can be present in significant amount in the ISM \citep[e.g., $\sim$30\% of the molecular gas in our Galaxy;][]{Grenier2005, Pineda2013}. 

One of the promising tracer of the diffuse gas is the OH molecule. Indeed it is found to be significantly widespread with respect to CO, and can be used to observe ISM phases that are partly atomic, partly molecular i.e., envelopes surrounding dense molecular clouds, which are neither detectable in CO nor HI emission. 
In fact, $\rm OH$ was the first molecule detected at radio wavelengths \citep{Weinreb1963}.  Since then, along with HCO$^+$ it has emerged as one of the best tracers of H$_2$ gas \citep[][]{Liszt1999}.
It is quite efficient to observe it through the four 18-$\rm cm$ ground-state $\Lambda$-doubling transitions which occur at rest frequencies of 1612.231, 1665.402, 1667.359 and 1720.530 MHz. The relative strengths of these lines in the local thermodynamic equilibrium is 1612:1665:1667:1720 MHz = 1:5:9:1.  But the line ratios are seldom found to be in LTE and are sensitive to the local physical conditions in the ISM.

In the Galaxy, $\rm OH$ is routinely detected in a broad range of astrophysical environments. This includes OH maser emission arising from the star-forming regions \citep{Caswell1999}, the envelopes of the late-type stars \citep{teLintelHekkert1989, Engels2015}, the proto-planetary nebulae \citep{teLintelHekkert1996} and the supernovae remnants \citep{Brogan2013}.  
The non-amplified $\rm OH$ emission and $\rm OH$ absorption are detected in the diffuse molecular clouds in the Galactic plane \cite[e.g.][]{Dawson2014}, the individual Giant Molecular Cloud complexes \citep[as W43][]{Walsh2016}, the clouds at the Galactic center \citep{Boyce1994}, the high-latitude diffuse and translucent clouds \citep{Grossmann1990, Cotten2012, Donate2019} and cirrus clouds of our Galaxy \citep{Barriault2010}. 
Notably, \citet{Li2018} recently published $\rm OH$ absorption line measurements from the Millenium survey \citep[][]{Heiles2003}. They find that most of the detections are associated with the diffuse and translucent clouds, i.e., probing the so-called ``CO-dark" gas, and the peak of the log-normal function fitted to the excitation temperature distribution is 3.5\,K. The latter explains the general difficulty in detecting OH emission from diffuse clouds.

Outside the Galaxy, $\rm OH$ is detected in many luminous infrared galaxies locally and up to redshifts of 0.265 \citep{Darling2002, Fernandez2010}, not only as a megamaser emission, but also in absorption \citep[][]{Mcbride2015}. However, in normal star-forming galaxies, the detections are sparse. To date only four intervening $\rm OH$ absorbers are known at $z>0$ \citep[][]{Chengalur1999, Chengalur2003, Kanekar2005, Gupta2018}. 
Three of these originate from the dense molecular gas in gravitationally-lensed systems, that preselect the line of sights with high dust-extinction passing through the regions close to the center of galaxies. Only recently the OH main lines have been systematically searched in a sample of quasar sight lines tracing the diffuse cold atomic gas \citep[][]{Gupta2018}. This led to the first successful extragalactic detection of $\rm OH$ originating from the diffuse gas \citep[see also][]{Combes2019}. 
The upcoming large blind \HI\ 21-cm absorption line surveys, particularly the MeerKAT Absorption Line Survey \citep[MALS; ][]{Gupta16} and the  the  First  Large  Absorption  Survey  in  \HI\ \citep[FLASH;][]{Allison2020}  with the Square Kilometer Array (SKA) precursors and pathfinders will have the sensitivity and redshift paths to search OH 18-cm lines in a wide-range of environments in the Galaxy and distant galaxies at $0<z<2$. A summary of these surveys is provided in Table~1 of \citealt[][]{Gupta16}. Additionally, extragalactic $\rm OH$ may also be searched with 
Five-hundred-meter Aperture Spherical Telescope  \citep[FAST;][]{Zheng2020} and upgraded Giant Metrewave Radio Telescope \citep[uGMRT;][]{Gupta2020}. Since the diffuse gas has much higher cross-section than the dense gas, one may expect that the former will be preferentially represented in these surveys. 
It is therefore of great interest to obtain the expected detection rates and the physical conditions in diffuse gas that will be probed by these surveys.  This requires a detailed understanding of the production of OH and the influence of external factors on the populations of levels.

The production of OH molecule through the $\rm O$-bearing cycle has been studied in detail for different phases of ISM by many groups  \citep[e.g.,][]{Black1973, Herbst1973, Prasad1980, LePetit2002, Hollenbach2012, Bialy2015}.  
Most of the theoretical studies of $\rm OH$ production either assumed some abundances of $\rm H_2$ in the medium or calculated abundances of various species using complex chemical reaction network and radiative transfer modeling. A key component of the modeling is \HI/$\rm H_2$ transition, since it is crucial to determine $\rm H_2$ abundance, which is an important ingredient of the production cycle of $\rm O$-bearing molecules. Recently, \citet{Sternberg2014} and \citet[][]{Bialy2016} presented a simple analytical formalism to describe \HI/$\rm H_2$ transition in the cold medium. In \citet{Balashev2020}, we successfully applied this formalism to describe the production of the $\rm HD$ molecule in a semi-analytical manner. In this paper, we expand this semi-analytical formalism to describe the production cycle of $\rm O$-bearing molecules in diffuse ISM. We show that the galactic measurements by \citet{Li2018} can be well explained by our calculations and provide an estimate of the expected physical conditions in such diffuse molecular clouds. More specifically, we find that the measurements in high-latitude clouds, probing the cold ISM, with $\rm H_2$ column densities $\log N \gtrsim 19$ (where $N$ in cm$^{-2}$). This, together with recent constraints of the $\rm H_2$ column density distribution function by \citealt{Balashev2018}, allow us to estimate the expected detection rates of $\rm OH$ in the blind absorption line surveys.

This paper is organized as follows. In Section~\ref{sec:description} we describe the semi-analytical model to calculate the abundances of the $\rm O$-bearing molecules in the diffuse ISM. In  Section~\ref{sec:results} we present results from the calculation and  show that these are in a good agreement with the estimates from  sophisticated \Meudon\ numerical code. We also  examine how the relative abundances of $\rm OH$/\HI\ and  $\rm OH/H_2$ depend on the variation of the physical conditions in the medium. In Section~\ref{sec:discussion} we show that our prescription explains the observations in the Galaxy. Subsequently, we use it to estimate the column density distribution function of $\rm OH$ and hence the expected incidence rate of $\rm OH$ detection from the upcoming blind absorption line surveys with a particular focus on MALS. We summarize our results in Section~\ref{sec:summ}. Throughout the paper we assume a flat $\Lambda$CDM cosmology model with $\Omega_{\Lambda}=0.7$ and $H_0 = 68$\,km\,s$^{-1}$\,Mpc$^{-1}$.

\section{Analytical description}
\label{sec:description}
We consider the homogeneous and isothermal medium with following physical parameters that determine the chemical abundances: the total hydrogen number density $n_{\rm H}^{\rm tot}$ (in the following we will use $n_2=n_{\rm H}^{\rm tot} / 100\,\rm cm^{-3}$), kinetic temperature $T$ (in K), the metallicity $Z$ (relative to solar), the incidence UV field of strength $\chi$ \citep[in the units of Mathis field, ][]{Mathis1983} and  cosmic rays ionization rate (CRIR), $\zeta$ (in units of $10^{-17}$\,s$^{-1}$). For the latter we will implicitly assume that it is the primary ionization rate per hydrogen atom. 

In this work we consider diffuse ISM, with typical number densities $n\sim10-1000$\,cm$^{-3}$, exposed by relatively mild UV fields $\chi<100$ and CRIR values $\zeta<100$. We also explicitly study the dependence of the abundances of various species  on the metallicity, which we varied in the range: $Z\sim0.01-1$. We fixed the temperature to $T=100$\,K, i.e., we did not consider it as an additional parameter. This is reasonable because the relevant reaction rates have little temperature dependence in the range ($50-200$\,K) corresponding to the diffuse ISM \citep[][]{Balashev2019}. 
Further, in the following for a species denoted by $\rm X$ we will use $n_{\rm X}$, $x_{\rm X}$, $d_{\rm X}$ and $N_{\rm X}$ to describe the number density, relative abundance, depletion and column density, respectively. We will also explicitly denote atomic hydrogen by \HI\ and express all column densities in cm$^{-2}$.

The ionization potentials of C and O (metals that are relevant for this study) are 11.2 and 13.6\,eV, respectively.  Therefore, in the diffuse medium, they are  predominantly in the ionized and neutral form, respectively. 
Hence, their number densities 
can be defined as 
\begin{align}
    n_{\rm C^{+}} \approx n_{\rm C}^{\rm tot} &\equiv x_{\rm C} n^{\rm tot}_{\rm H} \approx 2.7\times 10^{-4} Z d_{\rm C} n^{\rm tot}_{\rm H}, \\
    n_{\rm O} \approx n_{\rm O}^{\rm tot} &\equiv x_{\rm O} n^{\rm tot}_{\rm H} = 4.9 \times 10^{-4} Z d_{\rm O} n^{\rm tot}_{\rm H},
\end{align}  
Further, we assume the number density of electrons, $n_{e} = n_{\rm H^+} + n_{\rm C^+}$. Since in the CNM, He and O are predominantly in the neutral form we will neglect their contributions to $n_e$. 

Following \citet[][]{Bialy2019} for the depletions $d_{\rm C}$ and $d_{\rm O}$ we use
\begin{equation}
d_{\rm X} = \left\{
\begin{aligned}
    &1 - \delta_{\rm X}   &Z \ge Z_0 \\
    &1 - \delta_{\rm X} (Z/Z_0)^2  &Z < Z_0, \\
\end{aligned}
\right.
\end{equation}
where $Z_0=0.2$ and $\delta_{\rm X}$ is the specified depletion at solar metallicity. We used $\delta_{\rm C}=0.53$ and $\delta_{\rm O}=0.41$, which recover typical ISM abundances of carbon and oxygen in our Galaxy. 

\begin{figure*}
 \includegraphics[width=1.\textwidth]{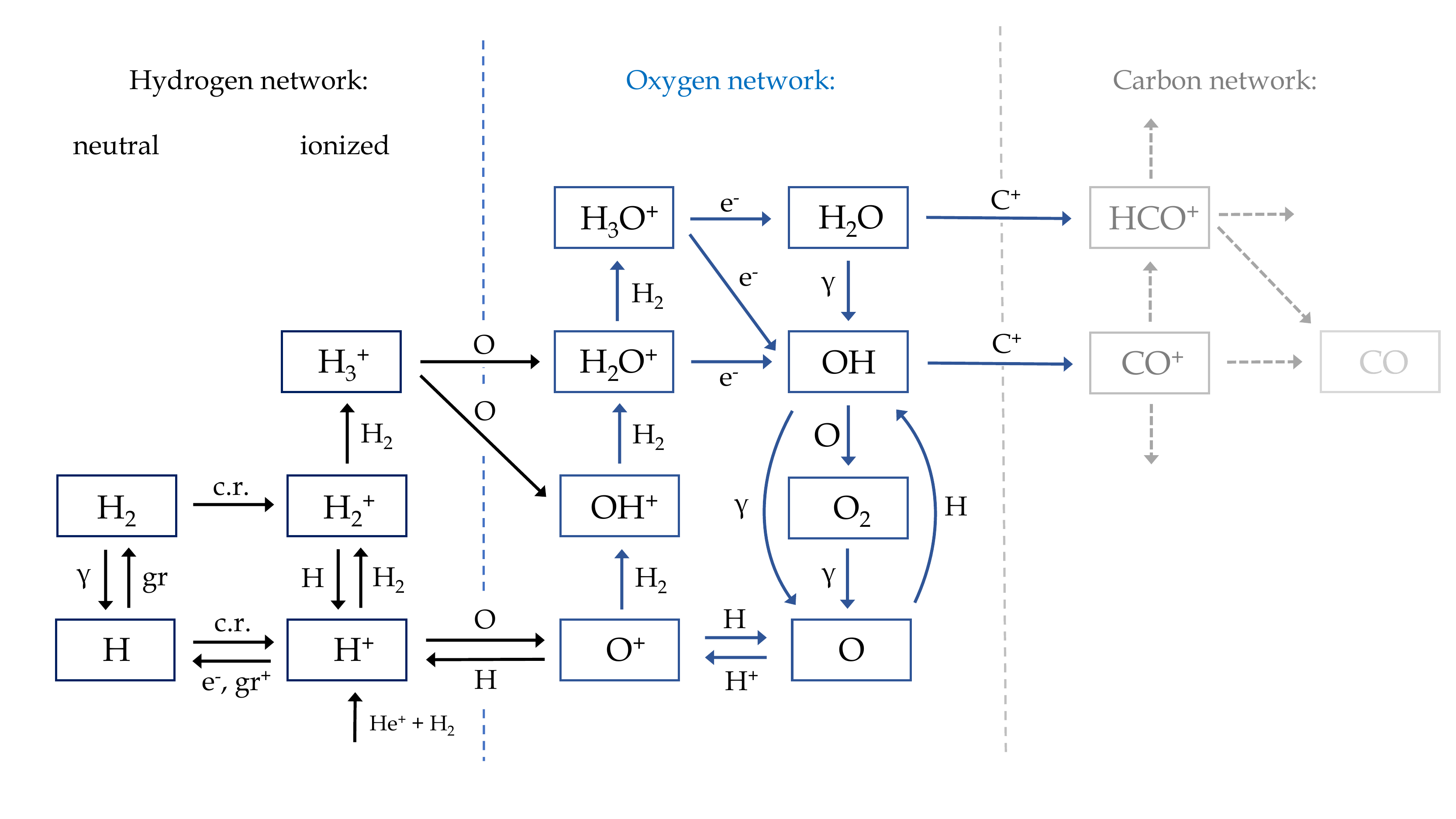}
 \caption{Simplified representation of chemical reaction network used in this paper. We plot only major reactions from the Table.~\ref{tab:reactions}. To keep the representation simple, we also do not show the channels that include $\rm He$ and $\rm Ar$ chemistry used in the model.  
 }
\label{fig:chem_networks}
\end{figure*}

In Table~\ref{tab:reactions}, we summarize reactions and their rates involved in the OH production. The focus is to write the (semi)-analytical description of the abundances of $\rm O$-bearing molecules. Therefore, we have restricted ourselves to the hydrogen and oxygen chemical networks, and consider only two reactions of $\rm O$-bearing species with $\rm C^+$, which are necessary to obtain correct destruction rates of $\rm OH$ and $\rm H_2O$ molecules in the diffuse ISM. In Figure~\ref{fig:chem_networks}, we present a simplified visual representation of the major reactions involved in the $\rm OH$ production cycle.  Such representations have been presented in the literature \citep[see, for examples,][]{Prasad1980, Bialy2015}. Most of the reaction rates have been taken from the UMIST database \citep{UMIST2012}. Note reaction \ref{eq:He^++H_c2} is missed in the  database. 
For $\rm O$-$\rm H$ charge exchange reactions, we used rates from \citet{Draine2011}.  Compared to UMIST, these provide a better fit to the detailed calculation by \citet{Stancil1999}. For the grain recombination reaction we used the fits to rates provided by \citet{Weingartner2001}. We did not study the grain and PAH charge distributions explicitly. 
For uniformity, the photo-dissociation rates and dust shielding have been taken from the recent extensive study by \citet{Heays2017}. These assume the plane-parallel geometry and the one-sided isotropic radiation field is of the spectral shape given by \citet[][]{Mathis1983}.

\begin{table*}
    \caption{Gas-phase reactions}
    \label{tab:reactions}
    \centering
    \abovedisplayskip=-10pt
    \belowdisplayskip=-10pt
    \begin{tabular}{m{4.5cm}cccccc}
    \hline
    \hline
    Reaction & Notation & $\alpha$ (cm$^{3}$\,s$^{-1}$) & $\beta$ & $\gamma$ & Refs & comment \\
    \hline
    {\begin{flalign}\rm H^+ + e \to H + h\nu \label{eq:H^++e}\end{flalign}} & $\alpha^{\rm rr}_{\rm H}$ & $3.5\times10^{-12}$ & $-0.7$ & .. & \citep{Prasad1980} & \\
    {\begin{flalign}\rm H^+ + O \to H + O^{+}\label{eq:H^++O}\end{flalign}} & $k_{\ref{eq:H^++O}}$ & $1.6 \times k^0_{\ref{eq:H+O^+}}$ & .. & 229 & \citep{Draine2011} & \\
    {\begin{flalign}\rm O^+ + H \to H^+ + O\label{eq:H+O^+}\end{flalign}}  & $k_{\ref{eq:H+O^+}}$ & \multicolumn{2}{c}{$k^0_{\ref{eq:H+O^+}} + k^1_{\ref{eq:H+O^+}} + k^2_{\ref{eq:H+O^+}}$ (see text)} & .. & \citep{Draine2011} & \\
    {\begin{flalign}\rm He^+ + e \to He + h\nu\label{eq:He^++e}\end{flalign}} & $\alpha^{\rm rr}_{\rm He}$ & $4.5\times10^{-12}$ & $-0.67$ & .. & \citep{Prasad1980} & \\
    {\begin{flalign}\rm He^+ + H_2 \to He + H + H^+\label{eq:He^++H_2_c1}\end{flalign}} & $k_{\ref{eq:He^++H_2_c1}}$ & $3.7\times10^{-14}$ & .. & $35$ & \citep{UMIST2012} & \\
    {\begin{flalign}\rm He^+ + H_2 \to He + H_2^+\label{eq:He^++H_2_c2}\end{flalign}} & $k_{\ref{eq:He^++H_2_c2}}$ & $7.2\times10^{-15}$ & .. & .. & \citep{UMIST2012} & \\
    {\begin{flalign}\rm He^+ + H \to He + H^+\label{eq:He^++H_c1}\end{flalign}} & $k_{\ref{eq:He^++H_c1}}$ & $1.2\times10^{-15}$ & $0.25$ & .. & \citep{Stancil1998} & \\
    {\begin{flalign}\rm He^+ + H \to HeH^+ + h\nu\label{eq:He^++H_c2}\end{flalign}} & $k_{\ref{eq:He^++H_c2}}$ & $1.7\times10^{-15}$ & $-0.37$ & .. & \citep{LePetit2006} & missed in UMIST \\
    {\begin{flalign}\rm HeH^+ + H \to He + H_2^+\label{eq:HeH^++H}\end{flalign}} & $k_{\ref{eq:HeH^++H}}$ & $8.2\times10^{-10}$ & $0.11$ & $31.5$ & \citep{Bovino2011} & \\
    {\begin{flalign}\rm HeH^+ + H_2 \to He + H_3^+\label{eq:HeH^++H_2}\end{flalign}} & $k_{\ref{eq:HeH^++H_2}}$ & $1.5\times10^{-9}$ & .. & .. & \citep{UMIST2012} & \\
    {\begin{flalign}\rm H_2^+ + H \to H_2 + H^+\label{eq:H_2^++H}\end{flalign}} & $k_{\ref{eq:H_2^++H}}$ & $6.4\times10^{-10}$ & .. & .. & \citep{Karpas1979} & \\
    {\begin{flalign}\rm H_2^+ + e \to H + H\label{eq:H_2^++e}\end{flalign}} & $k_{\ref{eq:H_2^++e}}$ & $1.6\times10^{-8}$ & $-1.18$ & 7.12 & \citep{EpeeEpee2016} & \\
    {\begin{flalign}\rm H_2^+ + H_2 \to H_3^+ + H\label{eq:H_2^++H_2}\end{flalign}} & $k_{\ref{eq:H_2^++H_2}}$ & $2.1\times10^{-9}$ & .. & .. & \citep{Theard1974} & \\
    {\begin{flalign}\rm Ar^+ + H_2 \to ArH^+ + H\label{eq:Ar^++H_2}\end{flalign}} & $k_{\ref{eq:Ar^++H_2}}$ & $8.4\times10^{-10}$ & $0.16$ & .. & \citep{Rebrion1989} & \\
    {\begin{flalign}\rm ArH^+ + e \to H + Ar\label{eq:ArH^++e}\end{flalign}} & $k_{\ref{eq:ArH^++e}}$  & $5\times10^{-10}$ & $-0.5$ & .. & \citep{Mitchell2005} & \\
    {\begin{flalign}\rm ArH^+ + O \to Ar + OH^{+}\label{eq:ArH^++O}\end{flalign}} & $k_{\ref{eq:ArH^++O}}$  & $8\times10^{-10}$ & .. & .. & \citep{Schilke2014} & \\
    {\begin{flalign}\rm ArH^+ + H_2 \to H_3^+ + Ar\label{eq:ArH^++H_2}\end{flalign}} & $k_{\ref{eq:ArH^++H_2}}$  & $8\times10^{-10}$ & .. & .. & \citep{Villinger1982} & \\
    {\begin{flalign}\rm H_3^+ + e \to H + H + H\label{eq:H_3^++e_c1}\end{flalign}} & $k_{\ref{eq:H_3^++e_c1}}$ & $4.4\times10^{-8}$ & $-0.52$ & .. & \citep{McCall2004} & \\
    {\begin{flalign}\rm H_3^+ + e \to H_2 + H\label{eq:H_3^++e_c2}\end{flalign}} & $k_{\ref{eq:H_3^++e_c2}}$ & $2.3\times10^{-8}$ & $-0.52$ & .. & \citep{McCall2004} & \\
    {\begin{flalign}\rm H_3^+ + O \to OH^+ + H_2\label{eq:H_3^++O_c1}\end{flalign}} & $k_{\ref{eq:H_3^++O_c1}}$ & $8.0\times10^{-10}$ & $-0.16$ & 1.4 & \citep{Bettens1999} & \\
    {\begin{flalign}\rm H_2 + O^+ \to OH^+ + H\label{eq:H_2+O^+}\end{flalign}} & $k_{\ref{eq:H_2+O^+}}$ & $1.7\times10^{-9}$ & .. & .. & \citep{Smith1978} & \\
    {\begin{flalign}\rm OH^+ + e \to H + O\label{eq:OH^++e}\end{flalign}} & $k_{\ref{eq:OH^++e}}$ & $3.75\times10^{-8}$ & $-0.50$ & .. & \citep{Mitchell1990} & \\
    {\begin{flalign}\rm OH^+ + H_2 \to H_2O^+ + H\label{eq:OH^++H_2}\end{flalign}} & $k_{\ref{eq:OH^++H_2}}$  & $1.0\times10^{-9}$ & .. & .. & \citep{Jones1981} & \\
    {\begin{flalign}\rm H_3^+ + O \to H_2O^+ + H\label{eq:H_3^++O_c2}\end{flalign}} & $k_{\ref{eq:H_3^++O_c2}}$ & $3.4\times10^{-10}$ & $-0.16$ & 1.4 & \citep{Bettens1999} & \\
    {\begin{flalign}\rm H_2O^+ + e \to O + H_2\label{eq:H_2O^++e_c1}\end{flalign}} & $k_{\ref{eq:H_2O^++e_c1}}$ & $3.9\times10^{-8}$ & $-0.50$ & .. & \citep{Rosen2000} & \\
    {\begin{flalign}\rm H_2O^+ + e \to O + H + H\label{eq:H_2O^++e_c2}\end{flalign}} & $k_{\ref{eq:H_2O^++e_c2}}$ & $3.1\times10^{-7}$ & $-0.50$ & .. & \citep{Rosen2000} & \\
    {\begin{flalign}\rm H_2O^+ + e \to OH + H\label{eq:H_2O^++e_c3}\end{flalign}} & $k_{\ref{eq:H_2O^++e_c3}}$ & $8.6\times10^{-8}$ & $-0.50$ & .. & \citep{Rosen2000} & \\
    {\begin{flalign}\rm H_2O^+ + H_2 \to H_3O^+ + H\label{eq:H_2O^++H_2}\end{flalign}} & $k_{\ref{eq:H_2O^++H_2}}$ & $6.4\times10^{-10}$ & .. & .. & 
    \citep{UMIST2012} & \\
    {\begin{flalign}\rm H_3O^+ + e \to OH + H_2\label{eq:H_3O^++e_c1}\end{flalign}} & $k_{\ref{eq:H_3O^++e_c1}}$ & $5.4\times10^{-8}$ & $-0.50$ & .. & \citep{Novotny2010} & \\
    {\begin{flalign}\rm H_3O^+ + e \to OH + H + H\label{eq:H_3O^++e_c2}\end{flalign}} & $k_{\ref{eq:H_3O^++e_c2}}$ & $3.1\times10^{-7}$ & $-0.50$ & .. & \citep{Novotny2010} & \\
    {\begin{flalign}\rm H_3O^+ + e \to H_2O + H\label{eq:H_3O^++e_c3}\end{flalign}} & $k_{\ref{eq:H_3O^++e_c3}}$ & $7.1\times10^{-8}$ & $-0.50$ & .. & \citep{Novotny2010} & \\
    {\begin{flalign}\rm H + O \to OH + h\nu\label{eq:H+O}\end{flalign}} & $k_{\ref{eq:H+O}}$ & $9.9\times10^{-19}$ & $-0.38$ & .. & \citep{LePetit2006} & \\
    {\begin{flalign}\rm OH + H \to H_2O + h\nu\label{eq:OH+H}\end{flalign}} & $k_{\ref{eq:OH+H}}$ & $4.0\times10^{-18}$ & $-2.00$ & .. & \citep{LePetit2006} & \\
    {\begin{flalign}\rm OH + H^+ \to H + OH^{+}\nu\label{eq:OH+H^+}\end{flalign}} & $k_{\ref{eq:OH+H^+}}$ & $2.1\times10^{-9}$ & $-0.50$ & .. & \citep{Prasad1980} & \\
    {\begin{flalign}\rm OH + C^+ \to CO^+ + H\label{eq:OH+C^+}\end{flalign}} & $k_{\ref{eq:OH+C^+}}$ & $7.7\times10^{-10}$ & $-0.50$ & .. & \citep{Prasad1980} & \\
    {\begin{flalign}\rm H_2O + C^+ \to HCO^+ + H\label{eq:H_2O+C^+_c1}\end{flalign}} & $k_{\ref{eq:H_2O+C^+_c1}}$ & $7.0\times10^{-10}$ & $-0.50$ & .. & \citep{Martinez2008} & \\
    {\begin{flalign}\rm H_2O + C^+ \to HOC^+ + H\label{eq:H_2O+C^+_c2}\end{flalign}} & $k_{\ref{eq:H_2O+C^+_c2}}$ & $1.4\times10^{-9}$ & $-0.50$ & .. & \citep{Martinez2008} & \\
    {\begin{flalign}\rm OH + O \to O_2 + H\label{eq:OH+O}\end{flalign}} & $k_{\ref{eq:OH+O}}$ & $3.7\times10^{-11}$ & $-0.25$ & $12.9$ & \citep{Prasad1980} & can be neglected \\
    \hline
    \multicolumn{5}{l}{Grain recombination:} \\
    {\begin{flalign}\rm H^+ + Gr^{+k} \to H + Gr^{k+1}\label{eq:H^++grain}\end{flalign}} & $\alpha^{\rm gr}_{\rm H}$ & \multicolumn{3}{c}{$\approx2.7\times10^{-10} \psi^{-0.9}$\,cm$^{-3}$\,s$^{-1}$} & \citep{Weingartner2001} & \\
    {\begin{flalign}\rm He^+ + Gr^{+k} \to He + Gr^{k+1}\label{eq:He^++grain}\end{flalign}} & $\alpha^{\rm gr}_{\rm He}$ & \multicolumn{3}{c}{$xxx$} & \citep{Weingartner2001} & \\
    \hline
    \multicolumn{6}{l}{Cosmic ray ionization} \\
    {\begin{flalign}\rm H + c.r \to H^+ + e\label{eq:H+cr}\end{flalign}} & $k_{\zeta}^{\rm H}$ & $1.7\times\zeta$  & .. & .. & \citep{Draine2011} & \\
    {\begin{flalign}\rm H_2 + c.r \to H^+ + H + e\label{eq:H_2+cr_c1}\end{flalign}} & $\tilde k_{\zeta}^{\rm H_2}$ & $0.17 \times \zeta$  & .. & .. & \citep{LePetit2002} & \\
    {\begin{flalign}\rm H_2 + c.r \to H_2^+ + e\label{eq:H_2+cr_c2}\end{flalign}} & $k_{\zeta}^{\rm H_2}$ & $3.4 \times \zeta$  & .. & .. & \citep{LePetit2002} & \\
    {\begin{flalign}\rm He + c.r \to He^+ + e\label{eq:He+cr}\end{flalign}} & $k_{\zeta}^{\rm He}$ & $3.4\times\zeta$  & .. & .. & \citep{LePetit2002} & \\
    {\begin{flalign}\rm Ar + c.r \to Ar^+ + e\label{eq:Ar+cr}\end{flalign}} & $k_{\zeta}^{\rm Ar}$ & $16.2\times\zeta$  & .. & .. & \citep{Schilke2014} & \\
    \hline
    \hline
    \multicolumn{6}{l}{Photodissociation reactions} \\
    Reaction & Notation & rate, $D_0$ (s$^{-1}$)\,$^{\rm c}$ & &  $\gamma_{\rm ds}$\,$^{\rm d}$ & Refs & comment \\
    \hline
    {\begin{flalign}\rm H_2 + h\nu\label{eq:H_2+photon}\end{flalign}} &  & $4.8\times10^{-11}$ & & 4.18 & \citep{Sternberg2014} & self-shielding important \\
    {\begin{flalign}\rm H_2^+ + h\nu\label{eq:H_2^++photon}\end{flalign}} & $D^{\rm H_2^+}$ & $3.9\times10^{-10}$ & & 2.78 & \citep{Heays2017} & can be neglected \\
    {\begin{flalign}\rm OH + h\nu\label{eq:OH+photon}\end{flalign}} & $D^{\rm OH}$ & $2.5\times10^{-10}$ & & 2.66 & \citep{Heays2017} &  \\
    {\begin{flalign}\rm OH^+ + h\nu\label{eq:OH^++photon}\end{flalign}} & $D^{\rm OH^+}$ & $1.1\times10^{-11}$ & & 3.97 & \citep{Heays2017} & can be neglected \\
    {\begin{flalign}\rm H_2O + h\nu\label{eq:H_2O+photon}\end{flalign}} & $D^{\rm H_2O}$ & $5.3\times10^{-10}$ & & 2.63 & \citep{Heays2017} &  \\
    {\begin{flalign}\rm ArH^+ + h\nu\label{eq:ArH^++photon}\end{flalign}} & $D^{\rm ArH^+}$ & $1.0\times10^{-11}$ & & 2.63 & \citep{Schilke2014} & can be neglected \\
    \hline
     \end{tabular}
    \begin{tablenotes}
    \item (a) The reaction rates are in the form $\alpha (T/300)^{\beta} e^{-\gamma/T}$.
    \item (b) References: (1) \citep{Draine2011}.
    \item (c) $D_0$ is a free-space photodissociation rate in Draine field \citep{Draine1996}, i.e., neglecting any extinction and self-shielding.
    \item (d) $\gamma_{\rm ds}$ is a constant in the factor related to the shielding by dust.  Following \citet{Heays2017}, $D = D_{0}e^{-\gamma_{\rm ds}A_V}$. 
    
    \end{tablenotes}
\end{table*}

\subsection{H$_2$ molecular fraction}
\label{sec:molecular_fraction}

Our description of the abundances of the species starts with defining the $\rm H_2$/\HI\ transition, which essentially governs the gross structure of the cloud. Following the formalism of \cite{Sternberg2014}, we express the H$_2$ molecular fraction $f_{\rm H_2}$ as a function of $N_{\rm H_2}$ through,
\begin{equation}
    \label{eq:f_H2}
    f_{\rm H_2} \equiv \frac{2n_{\rm H_2}}{n_{\rm H}^{\rm tot}} \equiv \frac{2n_{\rm H_2}}{n_{\rm H} + 2 n_{\rm H_2}} = \cfrac{2}{\alpha S_{\rm H2}(N_{\rm H_2}) e^{-\sigma_g (N_{\rm H} + 2 N_{\rm H_2})} + 2}, 
\end{equation}
where $S_{\rm H2}$ is the self-shielding function for H$_2$ \citep{Draine1996}, and $\sigma_g \approx 1.9\times 10^{-21} Z'\,\mbox{cm}^2$ is the dust LW-photon absorption cross-section per hydrogen atom. The ratio of free space H$_2$ photo-dissociation and H$_2$ formation on the dust grains, i.e., $\alpha$ is defined as,
\begin{equation}
\label{eq:alpha}
\alpha \equiv \frac{\chi D_0^{\rm H2}}{R^{\rm H2} n^{\rm tot}_{\rm H}} = 0.6\times 10^{4}n_2^{-1} \chi Z^{-1},
\end{equation}
where $D_0^{\rm H2}$ is the unattenuated photodissociation rate of H$_2$ in Mathis UV field and $R^{\rm H2}$ is the formation rate of H$_2$ on the dust grains. To be consistent with \Meudon\ models, we used $D_0^{\rm H2} = 4.8\times 10^{-11}$\,s$^{-1}$ and $R^{\rm H2} = 8 \times 10^{-17}\rm\,cm^{3}\,s^{-1}$ (see Sect.~\ref{sec:syst_RH2}). 
\citet{Sternberg2014} also found that $N_{\rm HI}$ 
can be expressed as a  function of $N_{\rm H_2}$ through, 
\begin{equation}
    \label{eq:HI_Sternberg}
    N^{\rm Stern}_{\rm HI} = \frac1{\sigma_g} \ln \left(\frac{\alpha G(N_{\rm H_2})}{2} + 1\right),
\end{equation}
where $G(N_{\rm H_2})$ is cloud-averaged H$_2$ self-shielding function. Here, we use the analytical fit to it from the paper by \citet{Bialy2016}.

In this formalism, the atomic hydrogen abundance drops to vanishingly small values in the $\rm H_2$ dominated core of the cloud, and hence the \HI\ column density has a fixed asymptotic value for each set of the $\alpha$ and $\sigma_g$ parameters. However, note that even at the cloud center,  due to dissociation of H$_2$ by cosmic rays, and/or desorption/adsorption reactions on the surface of dust grains the \HI  can be still present at reasonable levels. Based on the \Meudon\ calculations (see Section~\ref{sec:Meudon})  
we roughly estimated the $x_{\rm HI}$ to be $10^{-3} / Z$ in H$_2$-dominated regions. This gives the correction to the $N_{\rm HI}$ provided in equation~\eqref{eq:HI_Sternberg} as
\begin{equation}
\label{eq:NHI}
    N_{\rm HI} = N^{\rm Stern}_{\rm HI}(N_{\rm H_2}) + \frac{10^{-3}}{Z}N_{\rm H_2}.
\end{equation}

\subsection{$\rm O^+$, $\rm He^+$, $\rm HeH^+$ and $\rm ArH^+$}

The abundance of $\rm O^{+}$ is determined by the destruction with $\rm H_2$ \eqref{eq:H_2+O^+}, and $\rm O$-$\rm H$ charge exchange reactions \eqref{eq:H^++O} and \eqref{eq:H+O^+}, with rates $k_{\ref{eq:H^++O}} = 1.6 k_{\ref{eq:H+O^+}}^0 e^{-2.3/T_2}$ and $k_{\ref{eq:H+O^+}} = k_{\ref{eq:H+O^+}}^0 + k_{\ref{eq:H+O^+}}^1 + k_{\ref{eq:H+O^+}}^2$, where $k_{\ref{eq:H+O^+}}^0 = 2.6\times 10^{-10} T_2^{0.317 + 0.018\ln T_2}$\,cm$^3$s$^{-1}$, $k_{\ref{eq:H+O^+}}^1 = 9.3\times 10^{-11} T_2^{0.285 + 0.036\ln T_2}$\,cm$^3$s$^{-1}$, $k_{\ref{eq:H+O^+}}^2 = 1.5\times 10^{-10} T_2^{0.273 + 0.024\ln T_2} e^{-0.97/T_2}$\,cm$^3$s$^{-1}$, where $T_2 = T / (\rm 100\,K)$ \citep{Draine2011}. Using these reactions we get
\begin{equation}
    \label{eq:O^+}
    n_{\rm O^+} = \frac{k_{\ref{eq:H^++O}} n_{\rm H^+}n_{\rm O}}{k_{\ref{eq:H+O^+}} n_{\rm H} + k_{\ref{eq:H_2+O^+}} n_{\rm H_2}}.  
\end{equation}

The abundance of $\rm He^{+}$ is determined by cosmic ray ionization of He \eqref{eq:He+cr} and destruction in reactions~\eqref{eq:He^++grain}, \eqref{eq:He^++e}, \eqref{eq:He^++H_2_c1}, \eqref{eq:He^++H_2_c2}, \eqref{eq:He^++H_c1} and \eqref{eq:He^++H_c2}. Hence
\begin{equation}
    \label{eq:He^+}
    n_{\rm He^+} = \frac{k_{\zeta}^{\rm He} n_{\rm He}}{n_e \alpha^{\rm rr}_{\rm He} + \alpha^{\rm gr}_{\rm He}n_{\rm H}^{\rm tot} + (k_{\ref{eq:He^++H_2_c1}} + k_{\ref{eq:He^++H_2_c2}}) n_{\rm H_2} + (k_{\ref{eq:He^++H_c1}} + k_{\ref{eq:He^++H_c2}}) n_{\rm H}}.
\end{equation}

The $\rm He^+$ provides the main formation channel for $\rm HeH^+$, which is important for the formation $\rm H_2^+$ (in reaction~\eqref{eq:HeH^++H}) and $\rm H_3^+$ (in reaction~\eqref{eq:HeH^++H_2}). The abundance of $\rm HeH^+$ can be obtained as
\begin{equation}
    \label{eq:HeH^+}
    n_{\rm HeH^+} = \frac{k_{\ref{eq:He^++H_c2}} n_{\rm He^+} n_{\rm H}}{k_{\ref{eq:HeH^++H}} n_{\rm H} + k_{\ref{eq:HeH^++H_2}} n_{\rm H_2}}.
\end{equation}

The production of $\rm Ar^+$ is important for the $\rm H_3^+$ formation in diffuse atomic medium. $\rm Ar^+$ is formed in cosmic ray ionization and destructed mainly with $\rm H_2$ leading to production of $\rm ArH^+$, which in turn is destructed in reactions~\eqref{eq:ArH^++e} and \eqref{eq:ArH^++O} with $e$ and $\rm O$, respectively, and with $\rm H_2$ in reaction~\eqref{eq:ArH^++H_2}.  The latter results in $\rm H_3^+$ formation. Therefore we get
\begin{equation}
    \label{eq:ArH^+}
    n_{\rm ArH^+} = \frac{k_{\zeta}^{\rm Ar} n_{\rm Ar}}{k_{\ref{eq:ArH^++e}} n_{\rm e} + k_{\ref{eq:ArH^++O}} n_{\rm O} + k_{\ref{eq:ArH^++H_2}} n_{\rm H_2} + D^{\rm ArH^+}}.
\end{equation}


\subsection{H$_2^{+}$ and H$_3^+$}

The formation of $\rm H_2^+$ proceeds through two channels: one is due to cosmic rays ionization of $\rm H_2$ \eqref{eq:H_2+cr_c2}, and the other is due to $\rm HeH^+$ through reaction~\eqref{eq:HeH^++H}. Note, the reaction \eqref{eq:He^++H_2_c2} can be neglected. 
The destruction of $\rm H_2^+$ proceeds through reactions~\eqref{eq:H_2^++H}, \eqref{eq:H_2^++e}, \eqref{eq:H_2^++H_2} and photodissociation, which typically can be neglected. Hence,
\begin{equation}
    \label{eq:H_2^+}
    n_{\rm H_2^+} = \frac{n_{\rm H_2} k_{\zeta}^{\rm H_2} + k_{\ref{eq:HeH^++H}} n_{\rm HeH^+} n_{\rm H}}{k_{\ref{eq:H_2^++H}} n_{\rm H} + k_{\ref{eq:H_2^++e}}n_{e} + k_{\ref{eq:H_2^++H_2}} n_{\rm H_2} + D^{\rm H_2^+}}.
\end{equation}

In the medium where H$_2$ abundance is high, $\rm H_3^+$ molecule is preferentially formed from $\rm H_2^+$ through reaction~\eqref{eq:H_2^++H_2}, and in the atomic envelope of the cloud from $\rm ArH^+$ and $\rm HeH^+$ through  reactions~\eqref{eq:ArH^++H_2} and ~\eqref{eq:HeH^++H_2}, respectively. $\rm H_3^+$ is destroyed in reactions~\eqref{eq:H_3^++e_c1}, \eqref{eq:H_3^++e_c2},  \eqref{eq:H_3^++O_c1} and \eqref{eq:H_3^++O_c2}. Therefore
\begin{equation}
    \label{eq:H_3^+}
    n_{\rm H_3^+} = \frac{k_{\ref{eq:H_2^++H_2}} n_{\rm H_2^+} n_{\rm H_2} + k_{\ref{eq:ArH^++H_2}} n_{\rm ArH^+} n_{\rm H_2} + k_{\ref{eq:HeH^++H_2}} n_{\rm HeH^+} n_{\rm H_2}}{(k_{\ref{eq:H_3^++e_c1}}+k_{\ref{eq:H_3^++e_c2}}) n_{e} + (k_{\ref{eq:H_3^++O_c1}}+k_{\ref{eq:H_3^++O_c2}}) n_{\rm O}}.
\end{equation}

The chemistry of $\rm H_2^+$ and $\rm H_3^+$ was recently extensively discussed by \citet[][]{Neufeld2017} in the context of cosmic ray ionization rate in the diffuse medium. However, we note that to derive their equations they did not take into account the formation of $\rm H_2^+$ and $\rm H_3^+$ through $\rm HeH^+$ and $\rm ArH^+$, which resulted in smaller abundances of $\rm H_2^+$ and $\rm H_3^+$ in diffuse atomic part of the clouds, where $\rm H_2$ abundance is low.


\subsection{Ionization fraction}
\label{sec:ionization_fraction}

After determining $\rm He^+$ abundance, and knowing how to relate $\rm O^+$ and $\rm H^+$ abundance, the next step is to determine hydrogen ionization fraction. This is an important part of our  prescription to calculate the abundances of $\rm O$-bearing molecules. This is because the ionization fraction determines $\rm O^+$ abundance, and hence has a crucial importance on some of the channels for the production of $\rm O$-bearing molecules. We will show that once  ionization fraction is determined, we can write explicit analytical equations for abundances of $\rm O$-bearing molecules. 

The ionization fraction in the diffuse ISM, $f_{\rm H^+} = n_{\rm H^+} / n^{\rm tot}_{\rm H}$ can be obtained assuming that H$^{+}$  is mainly produced by ionization by cosmic rays (\ref{eq:H+cr} and \ref{eq:H_2+cr_c1}), charge exchange reaction with $\rm O^+$ \eqref{eq:H+O^+}, and from $\rm He^+$ and $\rm H_2^+$ in reactions \eqref{eq:He^++H_2_c1} and \eqref{eq:H_2^++H}, respectively. In turn $\rm H^+$ is destructed by radiative recombination \eqref{eq:H^++grain}, charge-exchange reaction between H$^{+}$ and oxygen \eqref{eq:H^++O}, and neutralization by grains \eqref{eq:H^++grain}. We neglect here the channels of the formation from $\rm D^+$ by reaction with $\rm H$ and $\rm H_2$ and destruction with $\rm D$, since they precisely equate each other, so they can be omitted from the balance equation for $\rm H^+$. Therefore
\begin{multline}
    \label{eq:H^+_balance_init}
    k_{\zeta}^{\rm H} n_{\rm H} + \tilde k_{\zeta}^{\rm H_2} n_{\rm H_2} + k_{\ref{eq:H+O^+}} n_{\rm O^+} n_{\rm H} + k_{\ref{eq:H_2^++H}} n_{\rm H_2^+} n_{\rm H} + k_{\ref{eq:He^++H_2_c1}} n_{\rm He^+} n_{\rm H_2} = \\ = \alpha^{rr} n_e n_{\rm H^{+}} + \alpha_{\rm H}^{\rm gr} n_{\rm H}^{\rm tot}n_{\rm H^{+}} + k_{\ref{eq:H^++O}} n_{\rm O} n_{\rm H^+}.
\end{multline}

From the balance equation \eqref{eq:O^+} for $n_{\rm O^+}$, we can write: $k_{\ref{eq:H^++O}} n_{\rm O} n_{\rm H^+} - k_{\ref{eq:H+O^+}} n_{\rm O^+} n_{\rm H} = k_{\ref{eq:H_2+O^+}} n_{\rm O^+} n_{\rm H_2}$.  Using this and the expressions \eqref{eq:O^+}, \eqref{eq:He^+} and \eqref{eq:H_2^+} for $n_{\rm O^+}$, $n_{\rm He^+}$ and $n_{\rm H_2^+}$, respectively, the equation \eqref{eq:H^+_balance_init} is modified to,
\begin{multline}
    \label{eq:H^+_balance}
    k_{\zeta}^{\rm H} n_{\rm H} + \tilde k_{\zeta}^{\rm H_2} n_{\rm H_2} + k_{\zeta}^{\rm H_2}  n_{\rm H_2} \frac{k_{\ref{eq:H_2^++H}}  n_{\rm H}  }{k_{\ref{eq:H_2^++H}} n_{\rm H} + k_{\ref{eq:H_2^++e}}n_{e} + k_{\ref{eq:H_2^++H_2}} n_{\rm H_2}} + \\ +  k_{\zeta}^{\rm He} n_{\rm He}\frac{k_{\ref{eq:He^++H_2_c1}} n_{\rm H_2}  +  \cfrac{k_{\ref{eq:He^++H_c2}} n_{\rm H} k_{\ref{eq:H_2^++H}}  n_{\rm H}}{k_{\ref{eq:H_2^++H}} n_{\rm H} + k_{\ref{eq:H_2^++e}}n_{e} + k_{\ref{eq:H_2^++H_2}} n_{\rm H_2}}\cfrac{k_{\ref{eq:HeH^++H}}n_{\rm H}}{k_{\ref{eq:HeH^++H}}n_{\rm H}+k_{\ref{eq:HeH^++H_2}}n_{\rm H_2}}}{n_e \alpha^{\rm rr}_{\rm He} + \alpha^{\rm gr}_{\rm He} n_{\rm H}^{\rm tot} + (k_{\ref{eq:He^++H_2_c1}} + k_{\ref{eq:He^++H_2_c2}}) n_{\rm H_2} + (k_{\ref{eq:He^++H_c1}} + k_{\ref{eq:He^++H_c2}}) n_{\rm H}}  \\ = (\alpha^{rr} n_{\rm e} +  \alpha_{\rm H}^{\rm gr} n^{\rm tot}_{\rm H}) n_{\rm H^{+}} +  \frac{k_{\ref{eq:H^++O}} k_{\ref{eq:H_2+O^+}} n_{\rm O} n_{\rm H_2}}{k_{\ref{eq:H+O^+}} n_{\rm H} + k_{\ref{eq:H_2+O^+}} n_{\rm H_2}} n_{\rm H^+}.
\end{multline}
This equation can be solved numerically, taking into account that the ion neutralization coefficients $\alpha_{\rm H}^{\rm gr}$ and $\alpha_{\rm He}^{\rm gr}$ are functions of the physical parameters. However, one can also write an approximate solution of this equation. Indeed, using \citet{Weingartner2001} we can express $\alpha_{\rm H}^{\rm gr}$ as,
\begin{multline}
\label{eq:alpha^gr}
    \alpha_{\rm H}^{\rm gr} = 2.7 \times 10^{-11} \psi^{-0.9} {\rm cm^{3}\,s^{-1}} \approx \\ \approx 1.6 \times 10^{-10} \chi^{-1} n_2 Z x_e {\,\,\rm cm^{3}\,s^{-1}} \equiv \tilde\alpha_{\rm H}^{\rm gr} x_e, 
\end{multline}
where $\psi = G_0 \sqrt{T/K} / (n_e / \rm cm^{-3})$ and $x_e = n_e / n^{\rm H}_{\rm tot} = x_{\rm C} + f_{\rm H^+}$. Here $G_0$ is the radiation field intensity relative to \citet{Habing1968}. This approximation, where we replaced power -0.9 by -1,  works well till  the following inequality, $\chi n_2^{-1} (10^{-4} / x_e) \gg 0.23$, valid.  This is indeed the case for physical parameter ranges relevant for the cold ISM.

Next, we neglect the $k_{\ref{eq:H_2^++e}} n_{e}$ and $\alpha^{\rm rr}_{\rm He}n_{e}$ terms in the left-hand part of the equation \eqref{eq:H^+_balance}. Then, substituting $n_{\rm H} = n_{\rm H}^{\rm tot} (1 - f_{\rm H_2})(1 - f_{\rm H^{+}})$, $n_{\rm H_2} = n_{\rm H}^{\rm tot} f_{\rm H_2}(1 - f_{\rm H^{+}})/2$, and $n_{\rm H^{+}} = f_{\rm H^{+}} n_{\rm H}^{\rm tot}$, and introducing
\begin{multline}
    \label{eq:subst}
    k_{\zeta}^{\rm eff} = k^{\rm H}_{\zeta} (1 - f_{\rm H_2}) + \tilde k^{\rm H_2}_{\zeta} f_{\rm H_2} / 2 + k_{\zeta}^{\rm H_2}  \frac{f_{\rm H_2}}{2} \left(1 +  \frac{k_{\ref{eq:H_2^++H_2}} f_{\rm H_2}}{2 k_{\ref{eq:H_2^++H}} (1 - f_{\rm H_2})}\right)^{-1} + \\ +  k_{\zeta}^{\rm He} f_{\rm He}\frac{k_{\ref{eq:He^++H_2_c1}} \cfrac{f_{\rm H_2}}{2}  + k_{\ref{eq:HeH^++H}} \left(1 +  \cfrac{k_{\ref{eq:H_2^++H_2}} f_{\rm H_2}}{2 k_{\ref{eq:H_2^++H}} (1 - f_{\rm H_2})}\right)^{-1}}{\alpha^{\rm gr}_{\rm He} + (k_{\ref{eq:He^++H_2_c1}} + k_{\ref{eq:He^++H_2_c2}}) f_{\rm H_2}/2 + (k_{\ref{eq:He^++H_c1}} + k_{\ref{eq:He^++H_c2}}) (1 - f_{\rm H_2})}
\end{multline}
we obtain the equation for $f_{\rm H^+}$ as
\begin{equation}
    \label{eq:x_HII_balance}
    f^2_{\rm H^+} + f_{\rm H^+} \left(x_{\rm C}  + \frac{k_{\ref{eq:H^++O}} x_{\rm O} A}{( \alpha^{rr} + \tilde\alpha^{gr})}  \right) = \frac{k_{\zeta}^{\rm eff}}{( \alpha^{rr} + \tilde\alpha^{gr}) n^{\rm tot}_{\rm H}}, 
\end{equation}
where $A = \left(1 + \frac{k_{\ref{eq:H+O^+}} n_{\rm H}}{k_{\ref{eq:H_2+O^+}}n_{\rm H_2}}\right)^{-1} \approx \frac{2 f_{\rm H_2}}{1 + f_{\rm H_2}}$ (at $T=100$\,K) and we take into account that  $(\alpha^{rr} +\ \tilde\alpha^{gr}) x_{\rm C} \gg k_{\zeta}^{\rm eff} / n^{\rm tot}_{\rm H}$. Note that a substitution similar to \eqref{eq:subst} was used by \citet{Glassgold1974}.

Based on the quadratic equation \eqref{eq:x_HII_balance} we can write the solution for $f_{\rm H^+}$  as
\begin{equation}
    \label{eq:x_HII}
    f_{\rm H^{+}} = \tilde x\left(\sqrt{\frac{k_{\zeta}^{\rm eff}}{(\tilde\alpha^{gr} + \alpha^{rr}) n^{\rm tot}_{\rm H} \tilde x^2} + 1} -1\right)
\end{equation}
where
\begin{equation}
    \tilde x = \frac12\left(x_{\rm C} + \frac{k_{\ref{eq:H+O^+}} x_{\rm O} A}{\alpha^{rr} + \tilde\alpha^{gr}}\right) 
\end{equation}

\subsection{$\rm OH^+$, $\rm H_2O^+$ and $\rm H_3O^+$}
 $\rm H_3^+$ and $\rm H_2^+$ molecules provide the main channels of $\rm OH^+$ formation following reactions~\eqref{eq:H_3^++O_c1} and \eqref{eq:H_2+O^+}. Once $\rm OH^+$ is formed it mainly produces $\rm H_2O^+$ in reaction~\eqref{eq:OH^++H_2}. Additionally, $\rm OH^+$ can be destructed in recombination with electrons \eqref{eq:OH^++e} and photodissociated \eqref{eq:OH^++photon}. Therefore
\begin{equation}
    \label{eq:OH^+}
    n_{\rm OH^+} = \frac{k_{\ref{eq:H_3^++O_c1}} n_{\rm H_3^+} n_{\rm O} + k_{\ref{eq:H_2+O^+}} n_{\rm O^+} n_{\rm H_2}}{k_{\ref{eq:OH^++H_2}} n_{\rm H_2}+ k_{\ref{eq:OH^++e}} n_{e} + D^{\rm OH^+}}.
\end{equation}

Apart from reaction~\eqref{eq:OH^++H_2}, there is an additional channel of $\rm H_2O^+$ formation in reaction~\eqref{eq:H_3^++O_c2}. The destruction $\rm H_2O^+$ includes production of $\rm H_3O^+$ in reaction~\eqref{eq:H_2O^++H_2} and three recombination channels with electrons~\eqref{eq:H_2O^++e_c1}, \eqref{eq:H_2O^++e_c2} and \eqref{eq:H_2O^++e_c3}. Hence 
\begin{equation}
    \label{eq:H2O^+}
    n_{\rm H_2O^+} = \frac{k_{\ref{eq:OH^++H_2}}n_{\rm OH^+} n_{\rm H_2} + k_{\ref{eq:H_3^++O_c2}} n_{\rm H_3^+}n_{\rm O}}{k_{\ref{eq:H_2O^++H_2}}n_{\rm H_2} + (k_{\ref{eq:H_2O^++e_c1}} + k_{\ref{eq:H_2O^++e_c2}} + k_{\ref{eq:H_2O^++e_c3}}) n_{\rm e}}.
\end{equation}
The abundance of H$_3$O$^{+}$ is determined by  
\begin{equation}
    \label{eq:H3O^+}
    n_{\rm H_3O^+} = \frac{k_{\ref{eq:H_2O^++H_2}}n_{\rm H_2O^+}n_{\rm H_2}}{(k_{\ref{eq:H_3O^++e_c1}} + k_{\ref{eq:H_3O^++e_c2}} + k_{\ref{eq:H_3O^++e_c3}}) n_{\rm e}}.
\end{equation}

Note that one of the channels of recombination of $\rm H_3O^+$ molecules leads to the formation of $\rm H_2O$ \eqref{eq:H_3O^++e_c3}, which can be photodissociated to $\rm OH$ or react with $\rm C^+$ to form $\rm HCO^+$ \eqref{eq:H_2O+C^+_c1} or its isomer $\rm HOC^+$ \eqref{eq:H_2O+C^+_c2}. 

\subsection{$\rm OH$ and $\rm H_2O$}
Finally, the number density of OH molecules, is determined by the formation in reactions~\eqref{eq:H_2O^++e_c3}, \eqref{eq:H_3O^++e_c1}, \eqref{eq:H_3O^++e_c2} and \eqref{eq:H+O}. Additionally, $\rm OH$ is formed by the photodissociation of $\rm H_2O$.  This, along with the destruction by reaction with $\rm C^+$ (see reactions~\eqref{eq:H_2O+C^+_c1} and \eqref{eq:H_2O+C^+_c2}), is the dominant destruction channel for $\rm H_2O$. 
$\rm H_2O$ is mainly produced by reaction~\eqref{eq:H_3O^++e_c3}. 
Therefore, we include the reaction~\eqref{eq:H_3O^++e_c3} among the  formation channels of $\rm OH$. The destruction of $\rm OH$ is due to reactions~ \eqref{eq:OH+H^+}, \eqref{eq:OH+C^+}, \eqref{eq:OH+O},   and photodissociation \eqref{eq:OH+photon}.  
The reaction \eqref{eq:OH+H^+} can dominate at low metallicities.

Using these reactions we can write that
\begin{multline}
    \label{eq:OH}
    n_{\rm OH} = \frac{\left(k_{\ref{eq:H_2O^++e_c3}}n_{\rm H_2O^+} + \left(k_{\ref{eq:H_3O^++e_c1}} + k_{\ref{eq:H_3O^++e_c2}}\right) n_{\rm H_3O^+}\right)n_{e} + k_{\ref{eq:H+O}} n_{\rm H} n_{\rm O}}{k_{\ref{eq:OH+C^+}}n_{\rm C^+} + k_{\ref{eq:OH+O}}n_{\rm O} + k_{\ref{eq:OH+H^+}}n_{\rm H^+} + D^{\rm OH}} +\\
    + \frac{k_{\ref{eq:H_3O^++e_c3}} n_{\rm H_3O^+}n_{e} }{k_{\ref{eq:OH+C^+}}n_{\rm C^+} + k_{\ref{eq:OH+O}}n_{\rm O} + k_{\ref{eq:OH+H^+}}n_{\rm H^+} + D^{\rm OH}} \frac{D^{\rm H_2O}}{(k_{\ref{eq:H_2O+C^+_c1}}+k_{\ref{eq:H_2O+C^+_c2}})n_{\rm C^+} + D^{\rm H_2O}}
\end{multline}

Finally, since $\rm OH$ and $\rm H_3O^+$ provide the main production channel of $\rm H_2O$ in the atomic envelope (where $\rm OH$ peredominantly formed in the  reaction \eqref{eq:H+O}) and molecular core, respectively, the $\rm H_2O$ abundance can be written using $\rm OH$ and $\rm H_3O^+$ abundances as
\begin{equation}
    \label{eq:H2O}
    n_{\rm H_2O} = \frac{k_{\ref{eq:OH+H}} n_{\rm OH} n_{\rm H} + k_{\ref{eq:H_3O^++e_c3}} n_{\rm H_3O^+} n_{\rm e}}{\left(k_{\ref{eq:H_2O+C^+_c1}}+k_{\ref{eq:H_2O+C^+_c2}}\right)n_{\rm C^+} + D^{\rm H_2O}}    
\end{equation}

\subsection{Asymptotics}
\label{sec:asymptotics}
Here, we obtain asymptotics for $\rm OH$ abundances in certain cases:
\begin{enumerate}
    \item Atomic gas. $f_{\rm H_2} \ll 1$.
    In this case, the OH is mainly produced in reaction~\eqref{eq:H+O} and destructed by photodissociation~\eqref{eq:OH+photon}. Hence  $x_{\rm OH}$ is simply 
    \begin{equation}
        x_{\rm OH} = \frac{n_{\rm OH}}{n_{\rm H}^{\rm tot}} = \frac{k_{\ref{eq:H+O}} n_{\rm O}}{D^{\rm OH} \chi} = 1.9 \times 10^{-10} \left(\frac{n_{2}Z}{\chi}\right)  d_{\rm O} \left(\frac{T}{300}\right)^{-0.38},
        \label{eq:x_OH_atomic}
    \end{equation}
    which gives $x_{\rm OH} \approx 1.2 \times 10^{-10} n_2 Z \chi^{-1}$ at T=100\,K and $d_{\rm O} =0.41$. It is evident that $x_{\rm OH}$ is proportional to the same combination of physical parameters $n_2 Z \chi^{-1}$ as $\rm H_2$ abundance in the atomic envelope, $n_{\rm H_2} = n_{\rm H}^{\rm tot} f_{\rm H_2} / 2 \approx n_{\rm H}^{\rm tot} / \alpha$ (see equations~\eqref{eq:f_H2}, \eqref{eq:alpha}). Therefore, the relative $\rm OH/H_2$ abundance will not be sensitive to the number density, metallicity and UV flux, and hence,
    \begin{equation}
        \frac{n_{\rm OH}}{n_{\rm H_2}} = \frac{\alpha k_{\ref{eq:H+O}} n_{\rm O}}{D^{\rm OH} \chi} = 2.5 \times 10^{-6} d_{\rm O}  \left(\frac{T}{300}\right)^{-0.38},
    \end{equation}
    which for $T=100$\,K and $d_{\rm O} =0.41$ is $1.5 \times10^{-6}$. 
    \item Molecular gas, $f_{\rm H_2} \lesssim 1$. 
    In this case the we can write,
    \begin{equation}
         x_{\rm OH} \approx x_{\rm O} \frac{k_{\ref{eq:H^++O}} n_{\rm H^+}}{D^{\rm OH}} \left(\frac{1 + \cfrac{k_{\ref{eq:H_2O^++e_c3}}n_e}{k_{\ref{eq:H_2O^++H_2}}n_{\rm H_2}}}{1 + \cfrac{k_{\ref{eq:H_2O^++e_c3}}n_e}{k_{\ref{eq:H_2O^++H_2}}n_{\rm H_2}}\cfrac{k_{\ref{eq:H_2O^++e_c1}} + k_{\ref{eq:H_2O^++e_c2}} + k_{\ref{eq:H_2O^++e_c3}}}{k_{\ref{eq:H_2O^++e_c3}}}}\right) \frac{1}{1 + \cfrac{k_{\ref{eq:H+O^+}}n_{\rm H}}{k_{\ref{eq:H_2+O^+}}n_{\rm H_2}}}   
    \end{equation}
    
    For  $T=100$\,K, we get
    \begin{equation}
    \label{eq:x_OH_molecular}
        x_{\rm OH} \approx  5.6\times  10^{-3}  \left(\frac{n_2 Z}{\chi}\right) d_{\rm O} \frac{1 + B}{1 + 5 B} \frac{2 f_{\rm H_2}}{1 + f_{\rm H_2}} f_{\rm H^+},
    \end{equation}
    where $B = \frac{k_{\ref{eq:H_2O^++e_c3}}n_e}{k_{\ref{eq:H_2O^++H_2}}n_{\rm H_2}} \approx 6 \times 10^{-2} \frac{Z}{f_{\rm H_2}}$, if the electrons are determined by the carbon abundance. Since $f_{\rm H^+}$ is a complex function of the physical parameters (see equation~\eqref{eq:x_HII}), it is hard to obtain a simple form of dependence on the physical parameters. 
    However, in the cold ISM $f_{\rm H^+}$ is typically in the range of $\sim 10^{-3} -10^{-5}$ 
    and it depends sub linearly on the physical parameters. Therefore,  as $f_{\rm H_2} \to 1$ we can write 
    \begin{equation}
    \label{eq:x_OH_molecular_high}
        x_{\rm OH} \approx  1.1\times  10^{-2} \left(\frac{n_2 Z}{\chi}\right) d_{\rm O} f_{\rm H^+}.
    \end{equation} 
    For intermediate $f_{\rm H_2}$ using equation~\eqref{eq:f_H2} this gives
    \begin{equation}
    \label{eq:x_OH_molecular_low}
        x_{\rm OH} \approx  7.2\times  10^{-7} \left(\frac{n_2 Z}{\chi}\right)^2 d_{\rm O} \frac{f_{\rm H^+}}{S_{\rm H_2}}.
    \end{equation}
    Thus, for intermediate $f_{\rm H_2}$, $x_{\rm OH}$ has a steeper dependence on $n_2Z\chi^{-1}$ .
    
\end{enumerate}

\section{Results}
\label{sec:results}
\begin{figure*}
    \centering
    \includegraphics[width=1.0\textwidth]{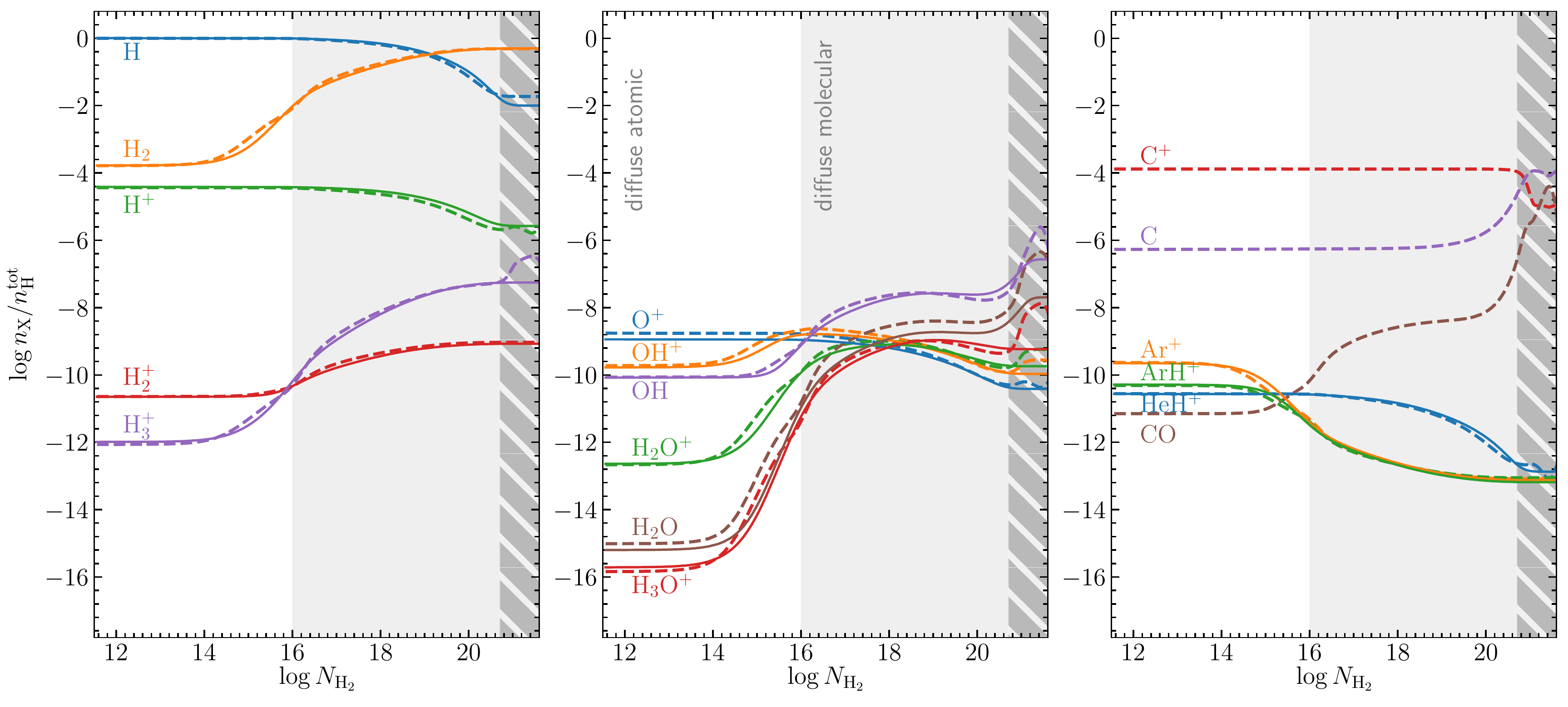}
    \caption{The relative abundances of different species involved in OH production as a function of $N_{\rm H_2}$ column density. The solid lines show the result of our calculation for a model with $Z=1$, $\chi=1$, $\zeta=3$, $n_{\rm H}^{\rm tot}=50$\,cm$^{-3}$, while the dashed lines represent \Meudon\ results. The filled gray region conditionally represents the diffuse molecular part of the cloud. Since we did not take into the the conversion of C$^+$ convert into C\,{\sc i} and CO and the carbon network reactions (involved mainly C and CO), therefore our calculations are not representative for the $\rm H_2$ column density, $\log N_{\rm H_2} \gtrsim 20.7$, at specified parameters of the model. This region is marked by hatch gray area. 
    }
    \label{fig:OH_Meudon}
\end{figure*}

In Fig.~\ref{fig:OH_Meudon}, we plot the abundances of various species estimated using the relationships discussed in Section~\ref{sec:description}. We consider $n_{\rm H}^{\rm tot}=50$, 
$\chi=1$, $\zeta=3$ and $Z=1$ 
i.e., the typical physical conditions in the local ISM. For the purpose of discussion of results,  we split the cloud into  three regions based on standard ISM classification: {\it (i)} diffuse atomic  -- where hydrogen is predominantly in atomic form; {\it (ii)} diffuse molecular -- where hydrogen is in molecular form but carbon is ionized; and {\it (iii)} dense molecular -- where hydrogen is in $\rm H_2$, and carbon is in neutral and molecular forms. 
We merge the dense molecular region with translucent one. Recall that our calculations in dense molecular region are not appropriate for the estimation of $\rm O$-bearing molecular abundances (see Sect.~\ref{sec:Meudon}). Indeed, we did not take into account $\rm C$ and $\rm CO$ chemistry that leads to significant drop in the electron fraction due to $\rm C^+/C$ transition. Additionally, once $\rm CO$ is formed it can cool the gas to temperatures $\sim 10$\,K, that can affect the reaction rates and number density.

For the diffuse atomic and molecular regions, our calculations reproduce the well-known behaviour of abundances of all the ions, molecules and radicals under consideration. Briefly, in the atomic envelope, the abundances of most molecular species, such as $\rm H_2^+$, $\rm H_3^+$, $\rm H_2O^+$ and $\rm H_3O^+$, are relatively low.  This is understandable because their production is coupled with the H$_2$ abundance. 
The main distinction is OH which can be formed through a neutral-neutral reaction \eqref{eq:H+O}, and the abundances of both $\rm H$ and $\rm O$ are relatively high in the atomic region. 
But, once the hydrogen starts to efficiently convert into the molecular form, due to self-shielding (at $\log N_{\rm H_2} \gtrsim 15$), there is a proportional increase in the abundances of aforementioned molecular species. 
Deeper into the cloud their abundances slightly reduce. This is due to the drop in the hydrogen ionization fraction, which leads to a drop in $\rm O^+$ and hence $\rm OH^{+}$ abundances. 
The latter provides a dominant channel for the production of $\rm H_2O^+$, $\rm H_3O^+$ and, consequently, $\rm OH$ in diffuse molecular clouds. 

The profiles of various species presented in Fig.~\ref{fig:OH_Meudon} would be vastly different for different physical conditions. For example, the low metallicity environment would result in much higher hydrogen ionization fraction \cite[see e.g.][]{Balashev2020} due to lower recombination rate on the dust grains, and also the electron fraction would be less sensitive to $\rm C^+$ abundance. Additionally, in case of extremely high CRIR, $\zeta \gtrsim 100$, the electron abundance will also be less sensitive to the $\rm C^+$ abundance, and $\rm H_2$ can be significantly dissociated by cosmic rays in the self-shielded regions of the cloud. 

\subsection{Comparison with \Meudon\ code}
\label{sec:Meudon}
Here, we compare our analytical calculations with the results from \Meudon\ code \citep{LePetit2006}. For \Meudon\ modelling, we consider  of slab gas irradiated from one side by the radiation field (with the Mathis spectrum) 
with constant density and temperature $T = 100$\,K.  
In general, we find that for the diffuse parts of the medium the results from our calculations for different species involved in OH production and OH itself are in a good agreement with \Meudon\ results. 

Fig.~\ref{fig:OH_Meudon} provides a comparison of our calculations and \Meudon\ results for a model with  $Z=1$, $\chi=1$, $\zeta=3$, $n=50$\,cm$^{-3}$.
For diffuse atomic and diffuse molecular regions of the cloud, the major differences between our model and \Meudon\ results are due to: 
{\it (i)} different reaction rates for the $\rm O$-$\rm H$ charge exchange reactions. This results in enhanced $\rm O^+$ abundance in \Meudon\ and hence an excess of $\rm O$-bearing molecules. (ii) Slightly different treatment of $\rm H_2$/\HI-transition. This results in  differences in the regions with $\log N_{\rm H2} \sim 14 - 16$ for $\rm H_2$, and hence, for all the other molecules and radicals. (iii) The difference in the $\rm H_3^+$, $\rm OH$, $\rm H_2O$ and $\rm H_3O^+$ photodissociation rates. Out of these, the most important is the difference for $\rm OH$, which is directly reflected in the difference in $\rm OH$ abundance. We verified that once we use \Meudon\ rates for the aforementioned reactions, there is almost one-to-one accordance between the results from our calculations and \Meudon\ code, with typical differences being $<30$ per cents. 

As previously mentioned, our model is not appropriate for the translucent parts of the cloud. But \Meudon\ profiles reproduce well the expected behaviour of molecular species discussed in previous section. 
However, here we are concerned only with the abundance of $\rm OH$ molecule in diffuse atomic and molecular regions.  Therefore, the results from our calculations are adequate for the purpose of this paper.


\subsection{Dependence on the physical conditions}

\begin{figure*}
 \includegraphics[width=1.\textwidth]{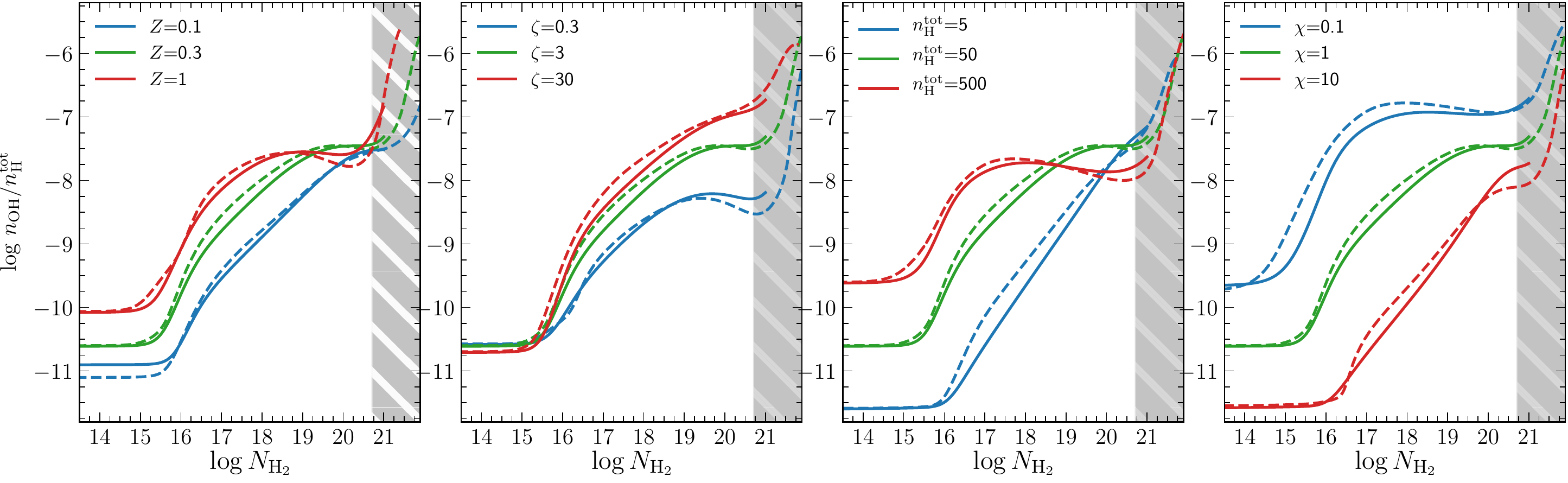}
 \caption{Dependence of $x_{\rm OH}=n_{\rm OH}/n^{\rm tot}_{\rm H}$  as a function of $N_{\rm H_2}$  on various physical parameters of the ISM. The green curves show the base model with $Z=0.3$, $\chi = 1$, $\zeta = 3$ and $n_{\rm H}^{\rm tot} = 100$. In each panel from left to right, the blue and red lines represent the profiles obtained by varying one of the physical parameters, i.e., metallicity, CRIR, UV field strength and number density (see legend in each panel), respectively.
 The solid and dashed lines correspond to our and \Meudon\ code calculation, respectively. The gray hatch portion schematically represents the translucent and dense molecular gas region, where our calculations are no longer applicable (the exact position of this region depends on the combination of the physical parameters and hence should be calculated individually for each line).
 }
\label{fig:xOH_H2_variation}
\end{figure*}

\begin{figure*}
 \includegraphics[width=1.\textwidth]{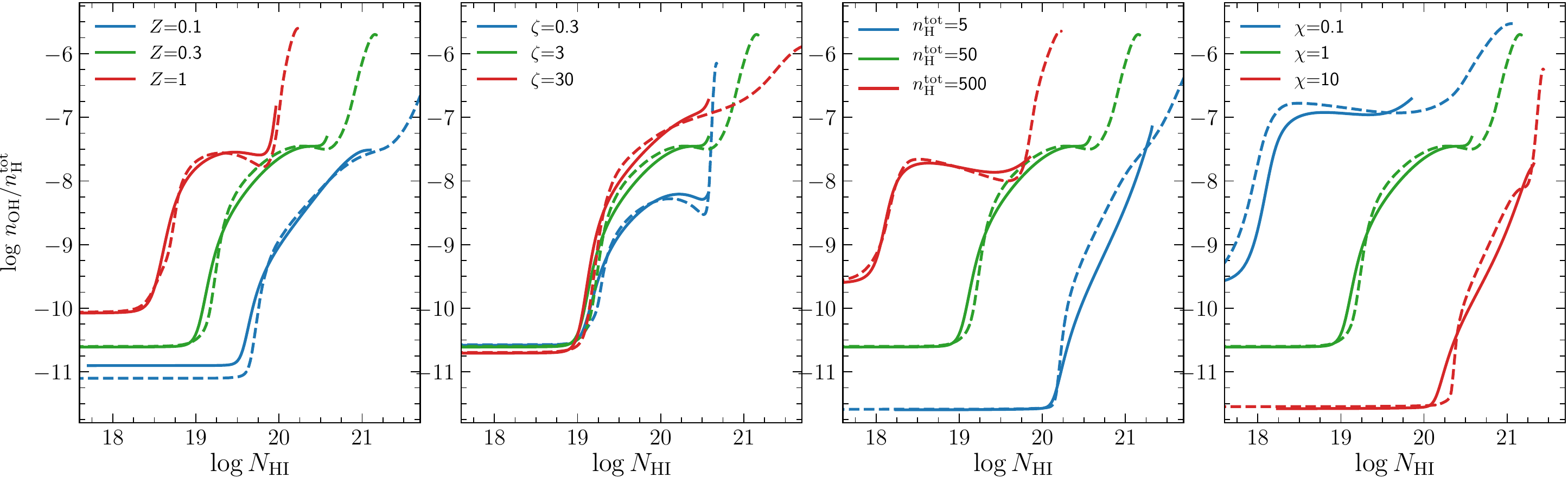}
 \caption{Dependence of $x_{\rm OH}=n_{\rm OH}/n^{\rm tot}_{\rm H}$ as a function of $N_{\rm HI}$ on various physical parameters of the ISM. The other details are the same as in Fig.~\ref{fig:xOH_H2_variation}. 
 }
\label{fig:xOH_HI_variation}
\end{figure*}

\begin{figure*}
 \includegraphics[width=1.\textwidth]{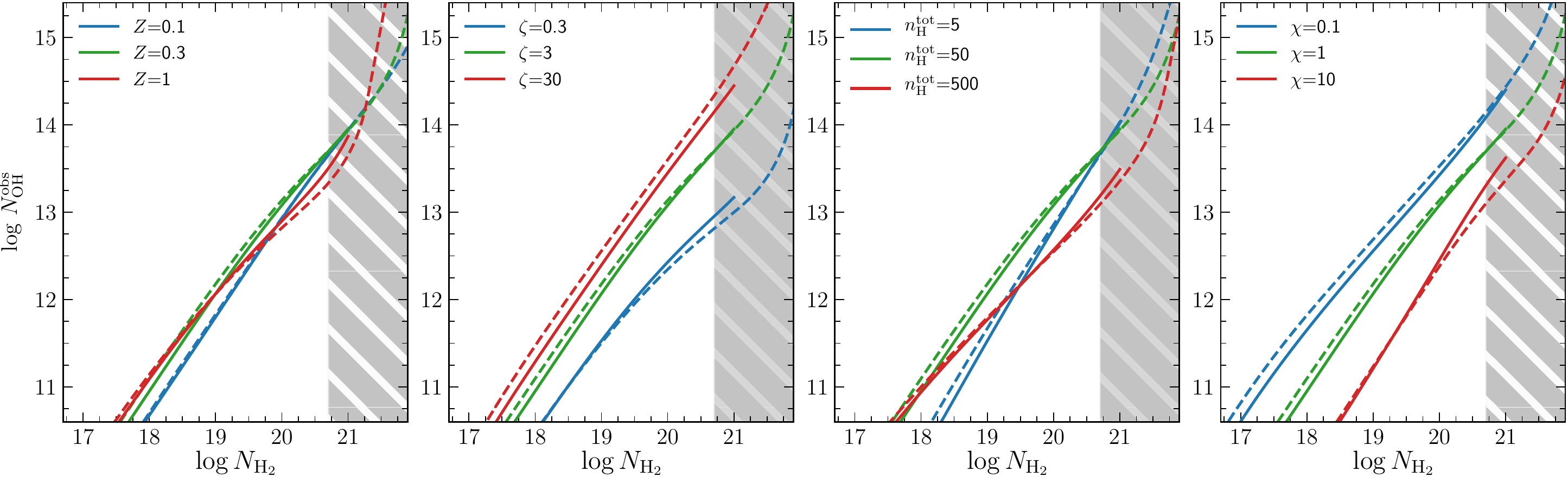}
 \caption{Dependence of $N_{\rm OH}$ as a function of $N_{\rm H_2}$  on various physical parameters of the ISM. The other details are the same as in Fig.~\ref{fig:xOH_H2_variation}. 
 }
\label{fig:NOH_H2_variation}
\end{figure*}

In Figs.~\ref{fig:xOH_H2_variation} and \ref{fig:xOH_HI_variation}, 
we present profiles of $x_{\rm OH}$ as a function of $N_{\rm H_2}$ and $\rm N_{\rm HI}$, respectively.  The profiles of $N_{\rm OH}$ versus $N_{\rm H_2}$ are presented in Fig.~\ref{fig:NOH_H2_variation}.
 Specifically, in each case we first obtained a base model which corresponds to $Z=0.3$, $\chi=1$, $\zeta=3$ and $n^{\rm tot}_{\rm H}=50$, and then independently varied metallicity, UV flux, CRIR,  and number density to cover a range of physical conditions in the medium. For completeness we also plot the profiles obtained using the \Meudon\ code. As expected, for diffuse medium we find a good agreement between our and \Meudon\ code calculations. Since the latter is able to correctly estimate $\rm OH$ abundances in translucent and molecular ISM, for completeness, we have retained calculations corresponding to these regions in the figures.

As obtained in Section~\ref{sec:asymptotics}, the $\rm OH$ abundance in the diffuse ISM directly depends on the combination $n_2 Z \chi^{-1}$. Such a dependence is clearly seen in Figs.~\ref{fig:xOH_H2_variation} and \ref{fig:xOH_HI_variation}. 
In the diffuse atomic ISM i.e., the regions with low $\rm H_2$ and low \HI\ column densities, this dependence is very stiff and $\rm OH$ abundance is well reproduced by the equation~\eqref{eq:x_OH_atomic}.
In the diffuse molecular region this dependence  gets altered as the ionization and molecular fractions, with latter also determining the $x_{\rm OH}$, are the functions of physical conditions $n_{\rm H}^{\rm tot}$, $\chi$, $\zeta$ and $Z$. 
Note that the dependence of $x_{\rm OH}$ on the cosmic ray ionization rate is not so strong. This is because the ionization fraction $f_{\rm H^+}$ depends sub-linearly on $\zeta$. However, in diffuse molecular ISM $x_{\rm OH}$ starts to have a linear dependence on $\zeta$.
Another interesting feature corresponds to deep into the cloud near the $\rm C^+/C/CO$ transition layer which demarcates the transition from the diffuse to dense gas phase. Here, for a wide range of physical conditions, $x_{\rm OH}$ tends to approach a constant asymptotic value which is reasonably reproduced by approximation given in equation~\eqref{eq:x_OH_molecular_high}. 


In Fig.~\ref{fig:NOH_H2_variation}, we plot how $N^{\rm obs}_{\rm OH}$ as a function of $N_{\rm H2}$ depends on the variation the physical conditions. We show $\rm OH$ column density as $N^{\rm obs}_{\rm OH} = 2 \times N_{\rm OH} (N_{\rm H_2}/2)$. 
Since our calculations are based on one-side radiation field models, this procedure simulates the slab medium exposed by radiation field and CRIR at both the sides. As discussed in subsequent sections, while comparing models with observations, we are interested only on column densities of $\log N_{\rm OH}\gtrsim12$, which corresponds to $\log N_{\rm H_2}\gtrsim 18$ (see Fig.~\ref{fig:NOH_H2_variation}).  
It is interesting to note that at these column densities, the relative $N_{\rm OH}/N_{\rm H_2}$ abundance is most sensitive to UV flux and CRIR (at $\zeta < 3$), but has very little sensitivity to metallicity. 
Additionally we note, that at the current observational sensitivity limits $N_{\rm OH} \gtrsim 10^{13}$\,cm$^{-2}$ the $\rm OH$ molecules probe the medium with $\rm H_2$ column densities $N_{\rm H_2} \gtrsim 10^{20}$\,cm$^{-2}$, that corresponding to the diffuse molecular ISM, where \HI/$\rm H_2$ transition is already complete.  From Fig.~\ref{fig:NOH_H2_variation} one can see  that the absorption systems with $N_{\rm OH}$ in ranges $ 10^{13}-10^{14}$\,cm$^{-2}$ correspond to the molecular gas before onset of $\rm CO$, i.e. the "CO-dark" molecular gas.

\subsection{Systematics and Limitations}

As already mentioned in previous sections our calculations are  applicable only to the diffuse atomic and molecular phases of ISM, i.e., well before the onset of $\rm C^+/C/CO$ transition. Here, we discuss other limitations and possible systematics that may impact the obtained profiles of the abundances of $\rm O$-bearing molecules which one should be aware of.

\subsubsection{Temperature gradient}
The $\rm O$-$\rm H$ charge exchange reactions are quite sensitive to the temperature of the gas. This is especially important if the gas gets heated to temperatures of $\sim 1000$\,K, for example, in the H$_2$ dissociated shell of an H$_2$ cloud. The situation corresponds to the case of a medium with low number densities ($n\lesssim {\rm few}\, \times\, 10$\,cm$^{-3}$) and exposed to high UV fluxes ($\chi\gtrsim 10$). In this case, in comparison to the isothermal case with $T\approx100$\,K the abundance of $\rm O^+$ is significantly enhanced.  At such high temperatures $\sim 1000$\,K, the $\rm H_2$ formation rate is also expected to be enhanced \citep[see e.g.][]{LeBourlot2012}, resulting in the enhancement of the $\rm O$-bearing molecules production in the atomic envelope of cloud. Both these factors result in the significant increase ($\gtrsim1$\,dex) for $\log N_{\rm OH}$. 
However, we note that such regions still have little effect on the total $\rm OH$ column density which are accessible through current observations ($\log N_{\rm OH}\gtrsim 12$). Additionally, these warm region are also expected to have high excitation temperatures so will not contribute significantly to the total  $\rm OH$ and \HI.  

\subsubsection{$\rm H_2$ formation rate}
\label{sec:syst_RH2}
The $\rm H_2$ formation rate, $R^{\rm H_2}$, is not well known in the ISM. The original measurements using {\sl Copernicus} satellite provided the first estimates of $R^{\rm H_2} \approx 3\times10^{-17}$\,cm$^3$s$^{-1}$ \citep{Jura1975}. The latter measurements updated this to slightly higher values $R^{\rm H_2} \approx 4\times10^{-17}$\,cm$^3$s$^{-1}$ \citep{Gry2002}. 
We use $R^{\rm H_2} \approx 8.0\times10^{-17}$\,cm$^3$s$^{-1}$ provided by the default calculations in \Meudon\ code with default 
dust size distribution parameters (power law with slope of $-3.5$ in the ranges of $3\times10^{-3}-3\times10^{-1}\,\mu$m) and mass to dust ratio of $6.3\times10^{-3}$, polycyclic aromatic hydrocarbons fraction of $4.6\times10^{-2}$. 

\subsubsection{$\rm H_2$ self-shielding function}

Since $\rm H_2$ self-shielding function determine the $\rm H_2$ abundance profile in the cloud, it has a direct impact on the $\rm OH$ abundance. In our calculations we used \citet{Draine1996} approximation, which is found to be in good agreement with the \Meudon\ code calculations which use the approximation given by \citet{Federman1979}. 
However, the self-shielding function depends on the H$_2$ level populations and raditaive transfer in H$_2$ lines.  Therefore, it 
is a complex function of the cloud structure. 
In the first approximation, the function significantly depends on the turbulent Doppler parameter in the medium. In our calculations, we used $b=2$\,km/s, which is the typical value observed in the cold diffuse ISM. We found that varying the doppler parameter in ranges $1..10$\, km/s resulted in insignificant changes of $N_{\rm OH}$ at any $N_{\rm H_2}$ in the observable ranges of $\log N_{\rm OH}\gtrsim12$. Note that this column density range already corresponds to the shielded medium. 

\subsubsection{{\rm OH} photodissociation rate}

The unshielded photodissociation rates used in our calculations are provided in Table~\ref{tab:reactions}. This implicitly assumes that the thickness of the medium is small, and the gas on the opposite side of the slab is not shielded. This is reasonable, since we are interested in the diffuse ISM, i.e., gas with small amounts of dust, which otherwise could be very effective in shielding.  
However, for some molecules, such as $\rm OH$, the shielding by $\rm H_2$ lines can be important, since the photodissociation cross-sections of $\rm OH$ and $\rm H_2$ are intersected. Therefore, locally this shielding will depend on the $\rm H_2$ column densities in the gas. 
Using the calculations by \citet{Heays2017}, one can estimate that when $\log N_{\rm H_2}$ is equal to 18 and 20 then the shielding of $\rm OH$ by H$_2$ is $\sim$ 0.95 and 0.68, respectively. Since, the abundance of $\rm OH$ scales directly  with UV flux, we can roughly estimate that taking into account this shielding will result in a $\lesssim 30\%$ increase of $\log N_{\rm OH}$ for the highest $\rm H_2$ column densities corresponding to the diffuse ISM. 

\subsubsection{Multilayer structure}
The observations of cold ISM always correspond to the measurement of quantities that are integrated over the line of sight (LOS). Therefore it is nearly impossible to distinguish the individual spatially-separated clouds with the same LOS velocities. The fact that the observed column densities at a given velocity can be a result of the concatenation of the several individual clouds, is refer to as the multilayer structure of the observed ISM. 
Indeed many studies provide evidences of the 
mutlilayer structure of the diffuse ISM \citep[e.g.][]{Bialy2017}. In this case, if the layers have the similar thickness in terms of $\rm H_2$, then the discrepancy in the calculations will be maximum.
Otherwise, the thicker component will likely dominate the observed column densities.


\subsubsection{Time dependent chemistry}
All calculations in our study were done with the assumption of the steady-state chemistry, i.e., when all the reaction rates are balanced. However, some reaction rates may have very long time scales in the diffuse ISM. 
The most important among these is the timescale for the formation of $\rm H_2$ on the dust grains, which can be estimated as $t_{\rm H_2}^{\rm form} \sim 1/ (R^{\rm H_2} n_{\rm H}^{\rm tot}) \sim 10 n_2$\,Myr. It is similar or even less than the dynamical timescales for some ISM clouds. 
A possible impact of the absence of steady-state is that H$_2$ abundance and hence $\rm OH$ abundance would be less than compared to the steady-state scenario. The overall impact of this on calculations is hard to properly quantify. The deviations from the steady-state are most likely induced by hydrodynamical motions in the medium. Hence, the time-dependent magneto-hydrodynamical and radiative transfer calculations with very high spatial resolution are required to properly resolve \HI-$\rm H_2$ transition and take into account ISM chemistry. Such modelling is hardly accessible with the state of art numerical codes. 
\section{Discussion}
\label{sec:discussion}
In this section, we compare the results from our calculations with observational data from our Galaxy and extragalactic OH absorbers. We also derive the incidence rate of OH absorbers, and discuss prospects with upcoming blind extragalactic \HI\ and OH absorption line surveys with a focus on MALS.

\subsection{Measurements in the Milky-Way}
\label{sec:measurements_mw}
\begin{figure*}
 \includegraphics[width=1.\textwidth]{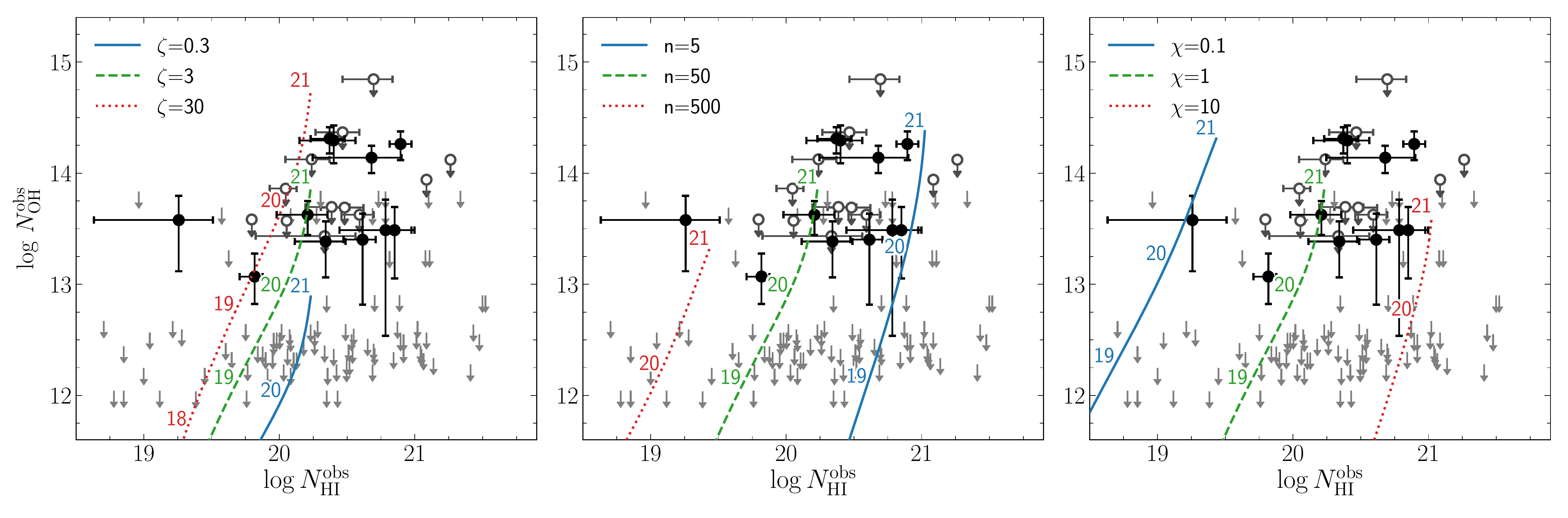}
 \caption{Comparison between relative $N_{\rm OH}$ and $N_{\rm HI}$ abundances from our calculations and observations of the Milky-Way \citep{Li2018}. The circles (filled and open) and gray arrows represent $\rm OH$ absorption detections and non-detections, respectively, for sightlines with confirmed \HI\ absorption detections. The error bars represent 1$\sigma$ uncertainties. 
 The green dashed curve in each panel represents the calculated $N_{\rm OH}$($N_{\rm HI}$) profile for the base model with $Z=1$, $\chi = 1$, $\zeta = 3$ and $n_{\rm H}^{\rm tot} = 50$\,cm$^{-3}$. The red dotted and blue solid curves show models obtained by varying a parameter with respect to the base model.  The parameter being varied i.e., cosmic ray ionization rate, UV field strength or number density is indicated in the legend of panels from left to right, respectively. 
 The numbers along the curves show corresponding $\log N_{\rm H_2}$.  
 }
\label{fig:NOH_HI_variation}
\end{figure*}
%
In Fig.~\ref{fig:NOH_HI_variation}, we compare the relative $N_{\rm OH}$ and $N_{\rm HI}$ abundances obtained using our calculations with the observations of the Milky-Way. 
The observational data are taken from the recent work by \citet{Li2018}, who presented data for \HI\ and OH  absorption from the Galaxy towards 44 extragalactic radio continuum sources. 
As can be seen from Fig.~8 of \citet{Li2018}, the errors on \HI\ and OH measurements in some cases are quite large.  
For our analysis, we have excluded {\it (i)} one data point which has no uncertainties on $N_{\rm OH}$, and {\it (ii)} treat all the $N_{\rm OH}$ measurements with errors extending down the y-axis in Fig.~8 of \citet{Li2018} as upper limits.  The latter are plotted as open circles in Fig.~\ref{fig:NOH_HI_variation}.  

The model profiles of $N_{\rm OH}(N_{\rm HI})$ from our calculations are shown as curves in Fig.~\ref{fig:NOH_HI_variation}.  
The $\rm OH$ and \HI\ column densities were calculated as $N^{\rm obs}_{\rm x} = 2 \times N_{\rm x} (N_{\rm H_2}/2)$ to simulate the situation where the slab medium is exposed by radiation field and CRIR at both sides.
In each panel, the base model with $Z=1$, $\zeta=3$, $n_{\rm H_2}^{\rm tot}=50$ and $\chi=1$ is shown as the dashed curve.  
Note that the relative $N_{\rm OH}$/$N_{\rm HI}$ abundance is mostly sensitive to UV field, and much less sensitive to CRIR. It is evident that the base model reproduces most of the observational data points reasonably well. However, some observed sightlines 
have a lower or higher relative $N_{\rm OH}$/$N_{\rm HI}$ abundance. The deviation of these points can be easily described assuming that the physical conditions have a dispersion within 1 dex of the base model.  
The majority of measurements with $N_{\rm OH}$ upper limits agree well with the model calculations.  The exceptions are line of sights with relatively high \HI\ but low $\rm OH$ column densities i.e., the bottom right corner in Fig.~\ref{fig:NOH_HI_variation}. 
Principally, these points can be explained by the concatenation of  several low-$N_{\rm HI}$ clouds along the line of sight. Since $N_{\rm OH}$ has a steep dependence on $N_{\rm HI}$ in region of $\log N_{\rm HI}~20-21$, a few components with high $N_{\rm HI}$ can even have relatively small $N_{\rm OH}$. 

\subsection{Measurements at $z>0$}
\label{sec:measurements_extra}
There are only four detections of $\rm OH$ absorption in intervening galaxies at $z>0$. Three of them are found towards strongly lensed radio sources: PKS\,$1830-211$ \citep[$z=0.886$][]{Chengalur1999}, B\,$0218+357$ \citep[$z=0.685$][]{Chengalur2003} and J\,$0134-0931$ \citep[$z=0.765$; ][]{Kanekar2005}. These sightlines  have $\rm OH$ column densities, $\log N_{\rm OH} \gtrsim 15$.  In the Galaxy, such column densities are associated with the dense molecular ISM. 
The absorber towards PKS\,$1830-211$ is particularly special.  More than fifty molecular species tracing dense gas corresponding to this have been detected \citep{wiklind1998, muller2011, muller2014}.
Our prescription is not applicable to dense gas phases associated with these.  

Here we focus on the remaining OH detection, which is at $z\approx0.05$ towards Q\,$0248+430$ \citep[][]{Gupta2018}. This sightline has $\log N_{\rm OH} \sim 14$ estimated using the excitation temperature, $T_{ex} = 3.5$\,K, which is comparable to the average Galactic measurements in the diffuse ISM. But the \HI\ column density, $\log N_{\rm HI} \sim 19.7$ estimated using the spin temperature, $T_{s}$ = 70\,K, is $\sim0.5$\,dex lower in comparison. 
A moderately high spin temperature, see for example \citet[][]{Gupta2018vlbi}, will reconcile the measurement with the base model and average Galactic measurements. 
Nevertheless, if that is not the case, then without any tension the measurement can be explained through a model with either relatively enhanced number densities $n_{\rm H}^{\rm tot} \gtrsim 500$ or CRIR $\zeta>100$ or a suppressed UV field $\chi<0.2$, or some other moderate combinations of these. 
In our opinion, the latter possibilities are plausible because the absorbing gas in this case is associated with the tidal tail from a galaxy at an impact parameter of $\sim 15$\,kpc.

Clearly, large samples of extragalactic \HI\ and $\rm OH$ measurements of diffuse ISM are required to disentangle various possibilities 
and make reasonable comparisons with the Galactic measurements. 

\subsection{OH column density distribution function}
\label{sec:cddf}

In this section we derive the expected $\rm OH$ column density distribution function which can be used to estimate the incidence rate of $\rm OH$ absorbers for blind absorption line surveys such as MALS. 
While investigating cosmological evolution of an absorber population, it is convenient to define the column density distribution function as \citep[see for example][]{Lanzetta1991}
\begin{equation}
    \cdd (N, X) = \frac{{\rm d}\mathcal{N}}{{\rm d}N{\rm d}X},
\end{equation}
where $N$ represents column density, $X$ represents absorption distance and ${\rm d}\mathcal{N}$ is the number of absorbers between $N$ and $N+{\rm d}N$, and $X$ and $X+{\rm d}X$. The use of absorption distance, $dX$, instead of the redshifts is convenient because it gives the value of distribution in comoving frame which guarantees that the objects with the same absorption cross-section will have the same column density function at all redshifts. Consequently, any cosmological evolution in the absorber population properties can be directly constrained.

As previously shown, at the column densities  relevant for observations ($\log N_{\rm OH}\gtrsim 12$), the OH molecule traces the gas with an associated H$_2$ column density of $\log N(\rm H_2) > 18$. At these column densities, we can use the measured H$_2$ column density distribution to determine the shape of the OH column density distribution function. Specifically, for OH we define
\begin{equation}
\begin{split}
{\cdd}_{\rm OH}(N_{\rm OH}, X) = \iiint \cdd_{\rm H_2}(N_{\rm H_2}, X) \frac{{\rm d} N_{\rm H_2}}{{\rm d} N_{\rm OH}} f_{\rm UV}(\chi) f_{n}(n) \\ f_{\rm cr}(\zeta) f_Z(Z) d\chi dn d\zeta dZ, 
\end{split}
\label{eq:FOH}
\end{equation}
where $N_{\rm H_2} \equiv N_{\rm H_2}(N_{\rm OH}, \chi, n, \zeta, Z)$. The functions $f_{\rm UV}(\chi)$, $f_{n}(n)$, $f_{\rm cr}(\zeta)$ and  $f_Z(Z)$ are the distributions expected for the UV field, number density, CRIR and metallicity, respectively. The integral needs to be performed over the each of these parameters. 
We can also use
\begin{equation}
\frac{{\rm d}N_{\rm H_2}}{{\rm d} N_{\rm OH}} = \frac{n_{\rm H_2}}{n_{\rm OH}}, 
\end{equation}
which ease the calculations, since $n_{\rm H_2}/n_{\rm OH}$ can be defined using Equations~\eqref{eq:OH} 
and \eqref{eq:f_H2}. 

It is reasonable to assume for $N(\rm H_2)$ column density distribution $\cdd_{\rm H_2}$ the shape of Schrecter function with the cut at the low column density region. This choice is motivated observationally. Firstly, at the high end of ${\rm H}_2$ column density distribution it can be well extrapolated from the measurements of the CO \citep{Zwaan2006}. Secondly, the direct measurements of the ${\rm H}_2$ column density column density distribution using the composite spectrum of the DLAs \citep{Balashev2018}, not only provides the constraints on the slope of the lower end of $\cdd_{\rm H_2}$, $\beta=-1.29 \pm 0.12$, but also shows that this slope is in a good agreement with the measurements based on the CO.
This value of the slope $\beta$ was recently confirmed in the statistical model for QSO-DLA absorbers \citep{Krogager2020}.
Additionally, the slope based on composite spectra is in the agreement with the estimated ${\rm H}_2$ column density distribution based on the detection in high-resolution spectra \citep{Noterdaeme2008}\footnote{However, current high resolution statistic may inherit observational bias, see discussion in  \citealt{Balashev2018}.} Hence for $\cdd_{\rm H_2}$ we use 

\begin{equation}
\cdd_{\rm H_2}(N, X) = C(z) N^{\beta} e^{-N / N^{\rm up}_{\rm H_2}} \Theta( N - N^{\rm low}_{\rm H_2} ),
\label{eq:FH2}
\end{equation}

where $C(z)$ is a scaling constant describing possible redshift evolution of column density distribution (i.e. we assume that the shape did not change). 
$\Theta$ is a Heaviside function, which take into account the abrupt drop in the number of the H$_2$ absorption systems at column densities lower. This drop of $\cdd_{\rm H_2}$ is connected with self-shielding of H$_2$ in the medium and we set $\log N^{\rm low}_{\rm H_2} = 18$ based both on observations and  sophisticated numerical simulations \citep[e.g.][]{Bellomi2020}. However, we note that the $\rm OH$ column density distribution at $\log N_{\rm OH}>13$ (corresponding to reasonable detection limit) will be little sensitive to the exact value of $N^{\rm low}_{\rm H_2}$ around chosen value, since at $\log N_{\rm H_2} \sim 18$ expected $\rm OH$ column densities, $\log N_{\rm OH} \lesssim 12$ for the wide range of the physical conditions (see Fig.~\ref{fig:NOH_H2_variation}).  Additionally, the calculations are valid only for the $\rm OH$ column densities $\log N(\rm OH) \lesssim 14$, since at higher $\rm OH$ column densities, $\rm OH$ is predominantly associated with dense CO-bearing medium. At this regime (corresponding to $\log N(\rm H_2) \gtrsim 21$), the $\rm OH$ column density at given $N_{\rm H_2}$ is significantly enhanced in comparison to extrapolation of our calculations (see Fig.~\ref{fig:NOH_H2_variation}). Therefore, our calculation doesn't sensitive to the exact choice of cutoff column density $N^{\rm up}_{\rm H_2}$ in equation~\ref{eq:FH2} unless $\log N^{\rm up}_{\rm H_2} \gtrsim 21$, which seems to be reasonable based on the local measurements \citep{Zwaan2006}, which found $\log N^{\rm up}_{\rm H_2} \approx 23.1$. We will keep later value in the following. Also one can note that typical cross-sections of the gas at the higher end of the $\rm H_2$ (and $\rm OH$) column density distribution is very low, and hence it will be technically difficult to probe this regime in observations.

For any redshift we can match the normalization constants of $\rm H_2$ and $\rm OH$ column density distributions
\begin{equation}
\int\limits_{10^{18}}^{\infty}{\mathcal{F}_{\rm H_2} {\rm d} N} = \frac{\rm d\mathcal{N_{\rm H_2}}}{{\rm d} X}(z) \approx \frac{\rm d\mathcal{N_{\rm OH}}}{{\rm d} X}(z) = \int\limits_{\approx10^{11}}^{\infty}{\mathcal{F}_{\rm OH} {\rm d} N}
\end{equation}
This equality is motivated by our modelling since we see that any sightline with $N_{\rm H_2} > 18$ will bear $\rm OH$ molecules with $N_{\rm OH} \gtrsim 11$ (see Fig.~\ref{fig:NOH_H2_variation}). At this lower end one can expect that the increase of the OH column density, comes from the CNM, where hydrogen is mostly in atomic form. However, the abundance of the $\rm OH$ is very low in such regions (see Equation~\eqref{eq:x_OH_atomic}) and therefore, to get $\log N_{\rm OH} \approx 11$ one should have $\log N_{\rm HI}\sim 21$ in the CNM. For such $\rm HI$ column densities it is expected that H$_2$/HI transition should already occur\footnote{This consideration do not take into account neither time-dependence of $\rm H_2$ formation in the CNM nor a possibility that large $\rm HI$ column density can be due to concatenation of the many layers of the CNM, each having  lower $\rm HI$ column density to have $\rm H_2$/$\rm HI$ transition.}. 
Additionally, this lower end of $\rm OH$ column density distribution is very challenging to assess at the sensitivity limits of the current facilities. 
From \citet{Balashev2018} normalization constant for H$_2$ column density distribution function, $\rm d\mathcal{N_{\rm H_2}}/{\rm d} X$, at $z\sim3$ is equal $\approx 10^{-2.4}$. Since there is no agreement on the evolution of $C(z)$ between the different modelling and observations, for the illustrations purposes in the following we will show derived quantities of OH column density distribution function and incidence rate scaled to the fraction of the diffuse molecular gas at $z\sim 3$ measured directly using QSO sightlines.

\begin{figure}
 \includegraphics[width=1.\columnwidth]{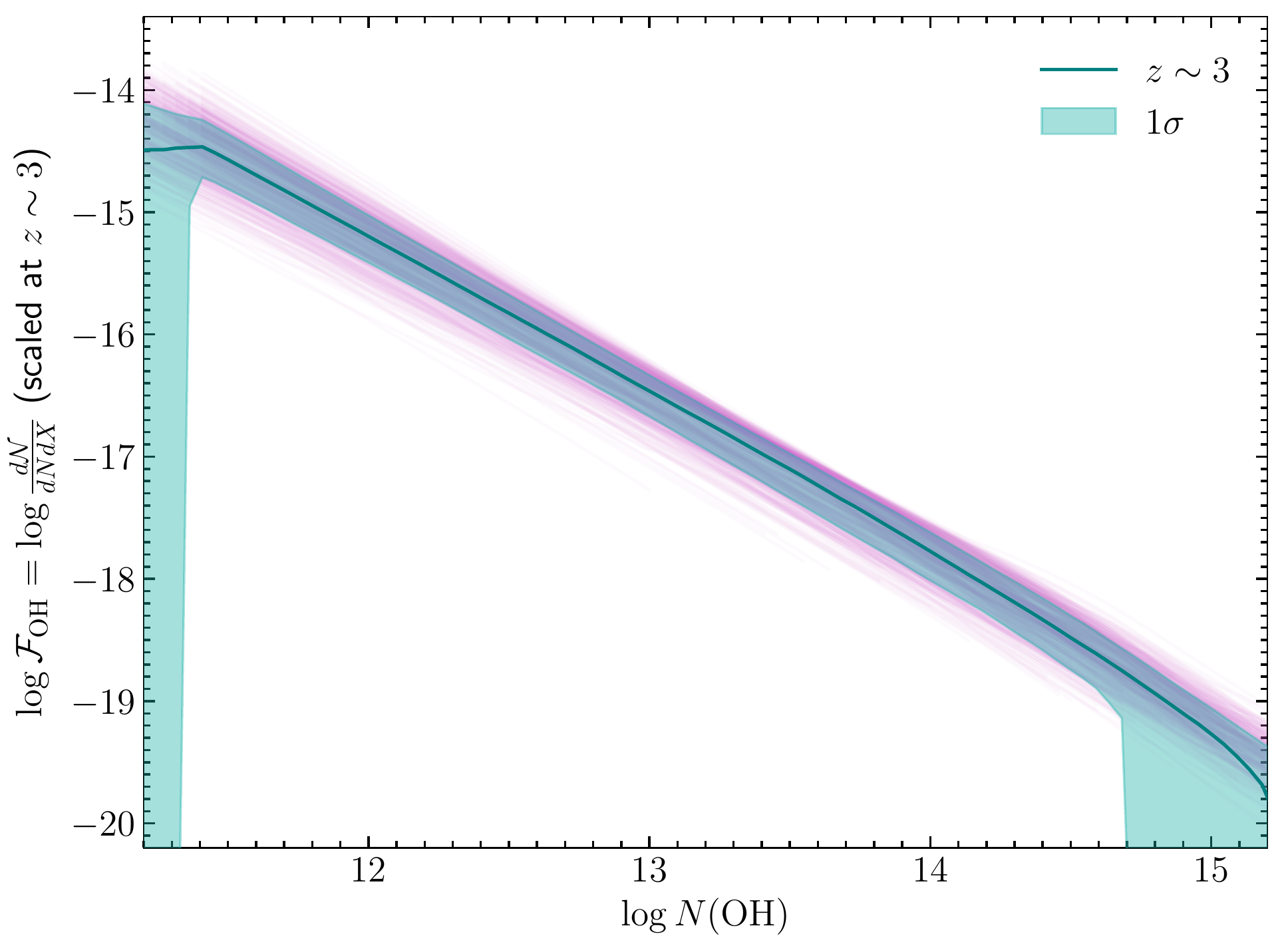}
 \caption{The OH column density distribution function, ${\mathcal F_{\rm OH}}$. The blue line show the median ${\mathcal F_{\rm OH}}$, while the blue region represents $0.68$ confidence region, based on the distribution of the individual ${\mathcal F_{\rm OH}}$ (shown by red lines), calculated for a range of physical conditions, with distributions determined by equation~\eqref{eq:f_p}.
 }
\label{fig:OH_cdd}
\end{figure}

To estimate $\rm OH$ column density distribution using equation~\eqref{eq:FOH} we assume that $f_{\rm UV}(\chi)$, $f_{n}(n)$, $f_{\rm cr}(\zeta)$, $f_Z(Z)$ are 
\begin{equation}
    \label{eq:f_p}
    f_{p} \sim 10^{\mu_p + \sigma_p \mathcal{N}(0,1)},
\end{equation}
where $\mathcal{N}(0,1)$ is a normal distribution with zero mean and dispersion equals one. The means and dispersion chosen to be ($\mu_{\chi} = 0, \sigma_{\chi}=0.5$), ($\mu_{n} = 2,\sigma_{n}=0.5$), ($\mu_{\zeta} = 0.5, \sigma_{\zeta}=0.5$) and ($\mu_{Z} = -0.5, \sigma_{Z}=0.5$), which correspond to the expected reasonable ranges. To calculate the $\rm OH$ column density distribution we randomly sampled parameters from these distributions. 
Then for each realization we also sampled the slope and normalization of the $\rm H_2$ column density distribution from the measurements using the Sloan Digitial Sky Survey (SDSS) composite spectrum \citep{Balashev2018} taking into account its anticorrelation. Using the dependence of the $N_{\rm H_2}(N_{\rm OH})$ this allow us to calculate the ${\mathcal F}_{\rm OH}$ for individual realization. 
The estimated median and dispersion of $\cdd_{\rm OH}$ is shown in the Fig.~\ref{fig:OH_cdd}. The $\rm OH$ column density distributions in the column density ranges $\log N_{\rm OH}=[12,15]$ have a power law dependence $\sim N^{\alpha}$ with $\alpha\sim[-1.2,-1.3]$. This is similar to slope of H$_2$ column density distribution $\beta=-1.29$, and can be easily justified, since there is the $N_{\rm H_2}/N_{\rm OH}$ factor in the Equation~\eqref{eq:FH2} has a almost constant at considered column densities and for the reasonable ranges of the physical parameters and hence $\alpha \approx \beta$. 

\subsection{The incidence rate and future prospects}
\label{sec:incid}

\begin{figure}
 \includegraphics[width=1.\columnwidth]{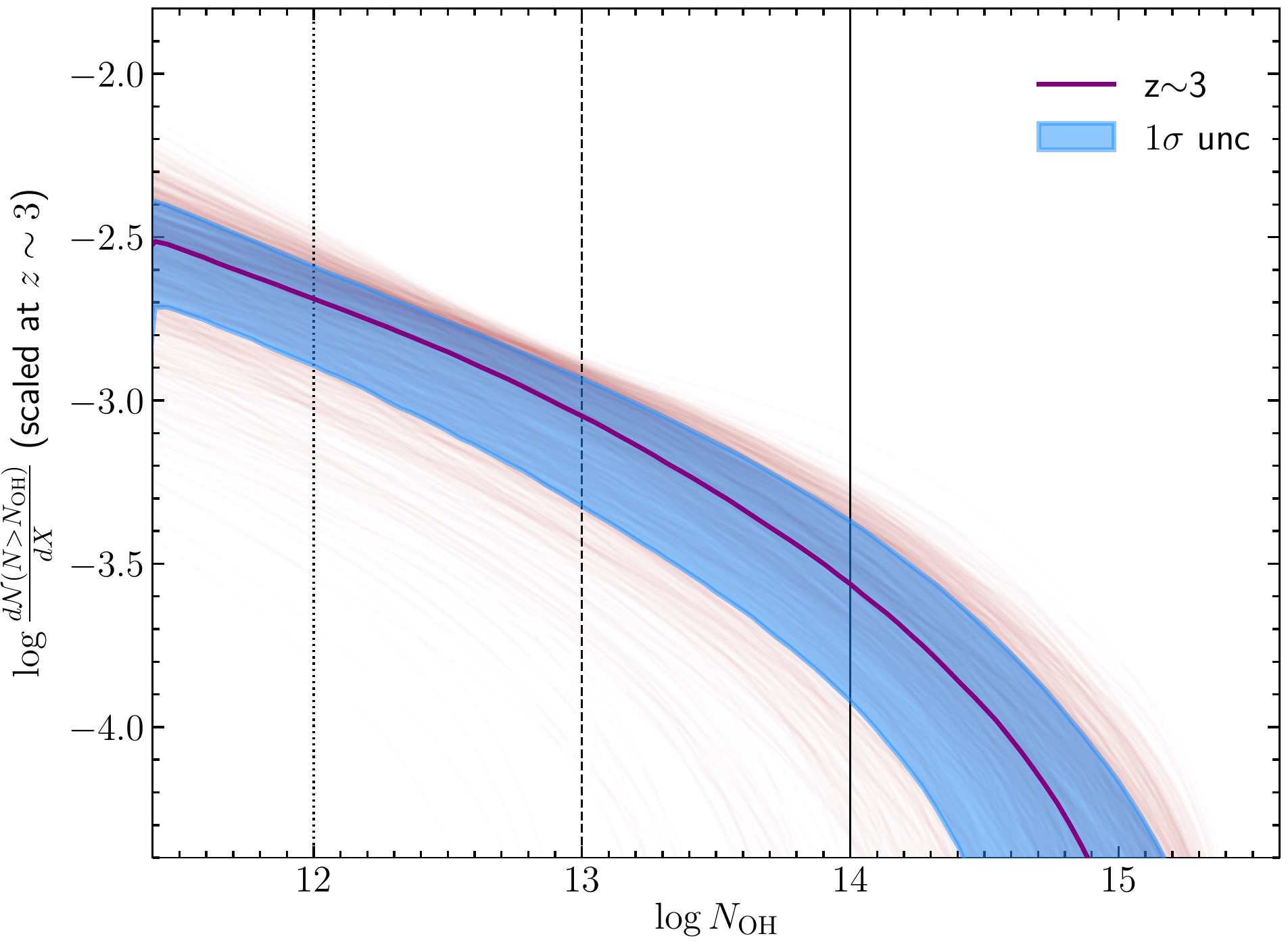}
 \caption{The incidence rate of absorption systems with $N > N_{\rm OH}$ as a function of $N_{\rm OH}$. The violet line indicate the averaged incidence rate, while the blue region represents $0.68$ confidence region, based on the distribution (shown by red lines), calculated for a range of expected physical conditions (see text).  
 }
\label{fig:OH_dndx}
\end{figure}

Using the derived OH column density distribution we estimate incidence rate i.e., the number of absorption systems per unit absorption distance with $N > N_{\rm OH}$ as a function of $N_{\rm OH}$. This quantity can be easily calculated as a survival function of the column density distribution function provided as
\begin{equation}
    \frac{{\rm d}\mathcal{N_{\rm OH}}(N>N_{\rm OH})}{{\rm d}X} = \int\limits_{N_{\rm OH}}^{\infty} \cdd_{\rm OH}(N, X) {\rm d} N
    \label{eq:OH_incidence}
\end{equation}
The estimated incidence rate of $\rm OH$ at $z\sim 3$ is shown in Fig.~\ref{fig:OH_dndx}. Based on this, one can see that at $z\sim 3$ to detect a single $\rm OH$ absorption at the sensitivity limits corresponding to $\log N_{\rm OH}$ of 12, 13 and 14, one needs to probe the total absorption distance paths, $\Delta X$, of  $480^{+270}_{-100}$, $1080^{+780}_{-230}$ and $3540^{+3870}_{-1140}$, respectively. 

Observationally, for an optically thin cloud, under the LTE conditions the integrated  optical depth of the strongest $\rm OH$ 18-cm line  is related to $N_{\rm OH}$ through,
\begin{equation}
N_{{\rm OH}}=2.24\times10^{14}~{T_{\rm ex}\over f_{\rm c}^{\rm OH}}\int~\tau_{1667}(v)~{\rm d}v~{\rm cm^{-2}}, 
\label{eqoh}
\end{equation}
where $T_{\rm ex}$ is the excitation temperature in Kelvin, $\tau_{1667}$($v$) is the optical depth of the 1667\,MHz line at 
velocity $v$, and $f_c^{\rm OH}$ is the covering factor \citep[e.g.,][]{Liszt1996}.
The milliarcsecond-scale spectroscopy of OH absorbers using the Very Long Baseline Interferometry (VLBI) is needed to probe the parsec-scale structures in the diffuse molecular gas and constrain $f_c^{\rm OH}$ \citep[][]{Srianand2013dib, Gupta2018vlbi}.
For the purpose of calculations presented in this paper, we assume that $f_c^{\rm OH}$ = 1 i.e., the absorbing gas completely covers the background radio continuum. 
Also, we adopt $T_{\rm ex}$ = 3.5\,K which is the peak of the log-normal function fitted to the $T_{ex}$ distribution of OH absorbers observed in the Galaxy \citep[][]{Li2018}. 
Note that if $T_{\rm ex}$ is coupled to the cosmic microwave background (CMB), then it will be higher at cosmologically significant redshifts.

Recently, \citet[][]{Gupta2020} used the uGMRT to carry out the first blind search of OH 18-cm main lines at $0.14 < z < 0.67$.  
For the modest total redshift path length of 24.4 corresponding to  $N_{\rm OH} > 2.4\times10^{14}(T_{\rm ex}/3.5)(1.0/f_{\rm c}^{\rm OH}) $\,\cmsq, they estimate the absorption per unit comoving path length to be, $l_{\rm OH}(z \sim 0.4) < 0.08$.
Due to the small absorption path length, not only the statistical power is limited but also the survey has probed only distant outskirts of star forming galaxies, mostly at impact parameters larger than 30\,kpc. 
Clearly, as motivated by this and the Fig.~\ref{fig:OH_dndx} much larger surveys are needed.

The next major improvement in the field of extragalctic OH absorption line search may come from MALS.  The survey is using MeerKAT's L- and UHF-bands covering 900 - 1670\,MHz and 580 - 1015\,MHz to search for OH 18-cm main lines at $0 < z < 0.85$ and $0.64 < z < 1.87$, respectively. The L-band observations of the survey are well underway. For the UHF-band the science verification observations are in progress. 
Each MALS pointing is centered at a radio source brighter than 200\,mJy at 1\,GHz.  The absorption lines will be searched towards this central  as well as numerous off-axis radio sources as faint as 5\,mJy with in the telescope's field-of-view \citep[FWHM $\sim 1.5^\circ$ at 1\,GHz;][]{Jonas2016}. Since the central radio source will be the brightest in the field-of-view, to maximize the low column density H~{\sc i} 21-cm absorption path of the survey, these are being selected to be typically at $z>0.6$ and $z>1.4$ for L- and UHF-bands, respectively.    

For MALS, a total of $\sim$1100 pointings, equally split among both the band, are expected to be observed.   
The target spectral rms is 0.5\,mJy\,beam$^{-1}$ per 5\,\kms.  
For the central bright sources, this implies a 5$\sigma$ intergated optical depth sensitivity, ${\cal{T}} (\equiv \int\tau dv) \le $ 0.066\,\kms.
This corresponds to a sensitivity to detect $N_{\rm OH} \ge  5\times 10^{13}$ (1.0/$f_c^{\rm OH}$)($T_{\rm ex}$/3.5)\,\cmsq\ towards all the sight lines.   
Considering the redshift distribution of MALS targets, the total absorption path lengths achieved towards the central bright radio sources for L- and UHF-band components of the survey are, $\Delta X \approx$ 600 and 1200, respectively. 
Based on our model predictions at $z\sim 3$, we get an incidence rate of $\approx4.1^{+1.9}_{-2.1}\times 10^{-4}$ for $\log N_{\rm OH}>13.7$. This roughly corresponds to $\Delta X\approx2440^{+2660}_{-780}$ needed to obtain a single detection.  
If this along with the 1$\sigma$ uncertainties presented in Fig.~\ref{fig:OH_dndx} is applied directly to the MALS search at $0 < z < 1.87$, only a handful of OH absorption detections are expected from the survey \citep[see also][for constraints based on Galactic measurement]{Gupta2018}.
But this extrapolation is highly uncertain on two accounts: {\it (i)} the $T_{ex}$ may be varying as a function of $z$, and {\it (ii)} the normalization at $z \sim 3$ i.e., $C(z)$ may be varying too.  

Here, it is worth elaborating on the following two caveats related to $C(z)$.  Recall that the $C(z)$ presented in Fig.~\ref{fig:OH_dndx} has been estimated using the $\rm H_2$ column density distribution at $z\sim 3$ measured using the UV absorption lines in the SDSS spectroscopic database.  
First, in SDSS the spectroscopic targets have been selected on the basis of optical colors. This makes the survey biased against dust-bearing sightlines. Since the presence of dust and cold gas ($\rm H_2$) are interlinked, the measurements of $\rm H_2$ column density distribution function based on SDSS should be treated as a lower limit.  
On the other hand, the off-axis radio sources in MALS which have not been included in the above estimate because they would generally be sensitive to $N_{\rm OH} > 14$, will have no optical or infrared color selection function applied.  Even the central MALS targets have been primarily selected on the basis of WISE infrared colors, and also include optically faint ($r >21$\,mag) targets which are generally excluded from optical spectroscopic surveys.  Therefore, MALS is much less biased against dust.
Second, caveat is related to the redshift evolution of $C(z)$ itself. 
Indeed, the simulations and emission line studies predict that the fraction of the molecular gas, $\Omega_{\rm H_2}$ (which is proportional to $C(z)$ in equation~\eqref{eq:FH2}), may show significant evolution, with $\Omega_{\rm H_2}$ several times higher than at $z=0$ and $z\sim 3$ \citep[see][and references therein]{Peroux2020}. 
These caveats may actually push $C(z)$ in a direction which is favorable for the upcoming lower redshift radio absorption lines surveys such as MALS.     
Indeed, \citet[][]{Muzahid2015} find the detection rate of $\rm H_2$ absorbers in DLAs and sub-DLAs at $z\sim0$ to be a factor of 2 higher than at $1.8<z<4.2$ \citep[][]{Noterdaeme2008}.

Eventually, the mid-frequency component of the upcoming first phase of the SKA, SKA1 (\textcolor{blue}{https://www.skatelescope.org/}), will have the frequency coverage (350 - 1760\,MHz) to search for OH main lines at $0<z<3.8$, and trace the cosmological evolution of OH abundance in diffuse ISM. 
The target column density sensitivity for a SKA1 survey ought to be $N_{\rm OH} \sim $ 12, and a total absorption line path in excess of 50,000.  Based on $C(z\sim 3)$, this will imply a prospect of detecting at least $\sim$100 OH absorbers and a realistic chance of constraining the column density distribution of diffuse OH.  The required integrated 5$\sigma$ optical depth sensitivity, ${\cal{T}}$ = 0.0013\,\kms\ is demanding.   Pragmatically such a survey will proceed commensally with other planned surveys \citep[see for example Table~1 of][]{Morganti2015}.   

It is important to bear in mind that currently the physical modelling and observations based on molecular emission lines can only constrain the evolution of the dense molecular gas. The direct correspondence of the diffuse and dense molecular gas in terms of the column density distribution is not at all constrained. 
In fact, the normalization $C(z)$ will depend on physical conditions i.e., $Z$, $n$, $\zeta$ and $\chi$ prevailing in the diffuse ISM at a certain redshift. Therefore, it may not scale with $\Omega_{\rm H_2}$ as simply stated above.   
Thus, the upcoming OH absorption line searches with the SKA  pathfinders and precursors, and eventually SKA1,  will be a powerful tool to constrain the physical conditions in the diffuse ISM and be highly complementary to ongoing molecular emission line surveys in millimeter regime that mostly probe the dense molecular gas.


\section{Conclusions}
\label{sec:summ}

In this paper, using a formalism based on the analytical description of  $\rm H I/H_2$ transition and a simplified network of major chemical reactions, we presented a semi-analytical prescription to estimate the abundances of O-bearing species in the diffuse ISM.  
We focused on $\rm OH$ molecule and our prescription is applicable only to diffuse ISM i.e., before $\rm C^+/C/CO$ transition, which corresponds to $N_{\rm OH} \lesssim 14$. 
We investigated the dependence of relative $\rm OH$/\HI\ and $\rm OH/H_2$ abundances on the variations of the physical conditions, i.e. the metallicity ($Z$), number density ($n$), cosmic ray ionization rate ($\zeta$) and strength of UV field ($\chi$) in the ISM.
For the reasonable ranges of the physical conditions in diffuse ISM (within $\pm1$\,dex of typical values) we showed that the abundances obtained using our simple prescription are in agreement with the calculation with the MEUDON PDR code which utilizes the full reaction network to predict abundances of various species. We confirmed that as was found previously, $\rm OH$ is strongly enhanced in the presence of $\rm H_2$.  We also confirmed that to the first order the $\rm OH$ abundance depends on the combination of $n Z/\chi$, while the scaling factor significantly increases as $\rm H_2$ molecular fraction approaches unity, resulting in typical $n_{\rm OH}/n^{\rm tot}_{\rm H}$ values of  $10^{-8}-10^{-7}$ in the $\rm H_2$-dominated region of the cloud. This  indicates that at current observational sensitivity limits  of $N_{\rm OH} \gtrsim 10^{13}$\,cm$^{-2}$, the $\rm OH$ molecules probe the medium with $\rm H_2$ column densities $N_{\rm H_2} \gtrsim 10^{20}$\,cm$^{-2}$, corresponding to the diffuse molecular ISM, where $\rm HI/H_2$ transition is already complete. Additionally, we show that the absorption systems with $N_{\rm OH}$ in ranges $ 10^{13}-10^{14}$\,cm$^{-2}$ probe molecular gas before onset of $\rm CO$, i.e. the "CO-dark" molecular gas.

We found that the Galactic measurements of OH main lines from \citet[][]{Li2018} can be reproduced by models with $n\sim 50$\,cm$^{-3}$, $\chi\sim 1$ (Mathis field) and $\zeta\sim3\times10^{-17}$\,s$^{-1}$, with a variation of about 1\,dex allowed around these values. The only available measurement of $\rm OH$ at $z>0$ in the diffuse regime ($\log N_{\rm OH} \lesssim14$) towards Q\,0248$+$430 from \citealt[][]{Gupta2018} indicates similar physical conditions.

Utilizing the observed $\rm H_2$ column density distribution function at $z\sim3$ \citep[][]{Balashev2018}, we derived the expected $\rm OH$ column density distribution function and incidence rate of $\rm OH$ absorbers. Both derived quantities include a scaling constant $C(z)$ which describes the possible redshift evolution of the cross-section of $\rm OH$-bearing diffuse molecular gas. We apply measured $C(z\sim3)$ directly to MALS \citep[][]{Gupta16}, the large survey project at the MeerKAT telescope, which will cover $0<z<1.87$ for the OH main lines.  
Considering only the central brightest source in the telescope's field-of-view, we estimate an $\rm OH$ incidence rate $dN/dX \approx4.1^{+1.9}_{-2.1}\times10^{-4}$ which indicates that MALS may only detect only a handful of $\rm OH$ absorbers from diffuse ISM.  
We discuss the caveats related to the extrapolation of $z\sim3$ distribution function to the lower redshift.  We suggest that the constraints on the $\rm OH$ based on this should be treated as a lower limit. The upcoming large radio absorption line surveys with SKA precursors and pathfinders will have the capacity to constrain $C(z)$ and inform the design of next generation surveys.

An ambitious SKA1 survey sensitive to detect $N_{\rm OH} \sim $ 12 and a total absorption line path in excess of 50,000, will have the prospect of detecting at least $\sim$100 OH absorbers and a realistic chance of constraining the column density distribution of diffuse OH.
This will be highly complementary to the molecular emission line surveys and physical models of ISM which are mostly focused on the dense molecular gas. 

\section*{Acknowledgements}

This work was supported by RSF grant 18-12-00301. 
SB and NG also thank the Munich Institute for Astro- and Particle Physics (MIAPP), which is funded by the Deutsche Forschungsgemeinschaft (DFG, German Research Foundation) under Germany's Excellence Strategy – EXC-2094 – 390783311, for the hospitality during the “Galaxy Evolution in a New Era of HI Surveys” workshop, where this work were initiated.


\bibliographystyle{mnras}
\bibliography{references.bib}

\label{lastpage}

\end{document}